\documentclass{aa}  

\DeclareUnicodeCharacter{FEFF}{}
\usepackage{natbib}
\usepackage[]{fontenc}
\usepackage{ae,aecompl}
\usepackage[dvipsnames]{xcolor}
\usepackage{psfrag,color}
\usepackage[]{inputenc}
\usepackage{graphicx}   
\usepackage{txfonts}    
\usepackage{amssymb}    
\usepackage{longtable}
\usepackage{subfigure}
\usepackage{adjustbox}

\def\la{\mathrel{\mathchoice {\vcenter{\offinterlineskip\halign{\hfil
$\displaystyle##$\hfil\cr<\cr\sim\cr}}}
{\vcenter{\offinterlineskip\halign{\hfil$\textstyle##$\hfil\cr<\cr\sim\cr}}}
{\vcenter{\offinterlineskip\halign{\hfil$\scriptstyle##$\hfil\cr<\cr\sim\cr}}}
{\vcenter{\offinterlineskip\halign{\hfil$\scriptscriptstyle##$\hfil\cr<\cr\sim
\cr}}}}}
\def\ga{\mathrel{\mathchoice {\vcenter{\offinterlineskip\halign{\hfil
$\displaystyle##$\hfil\cr>\cr\sim\cr}}}
{\vcenter{\offinterlineskip\halign{\hfil$\textstyle##$\hfil\cr>\cr\sim\cr}}}
{\vcenter{\offinterlineskip\halign{\hfil$\scriptstyle##$\hfil\cr>\cr\sim\cr}}}
{\vcenter{\offinterlineskip\halign{\hfil$\scriptscriptstyle##$\hfil\cr>\cr\sim\cr}}}}}
\begin{document} 

  \title{On the most luminous planetary nebulae of M31}

  \author{Rebeca Galera-Rosillo, 
        \inst{1,2,3}, Antonio Mampaso
        \inst{2,3}, Romano L.M. Corradi
        \inst{4,2}, Jorge Garc\'ia-Rojas
        \inst{2,3}, Bruce Balick
        \inst{5}, David Jones
        \inst{2,3}, Karen B. Kwitter
        \inst{6}, Laura Magrini
        \inst{7}, Eva Villaver
        \inst{8}}
\offprints{A. Mampaso} 

\institute{Isaac Newton Group of Telescopes, Apartado de Correos 321, Santa Cruz de La Palma, E-38700, Spain 
\and
Instituto de Astrof\'isica de Canarias, Calle V\'ia L\'actea, s/n, E-38205, La Laguna, Tenerife, Spain
\and
Departamento de Astrof\'isica, Universidad de La Laguna, E-38206, La Laguna, Tenerife, Spain           
\and
GRANTECAN, Cuesta de San José s/n, E-38712 , Breña Baja, La Palma, Spain
\and
Department of Astronomy, University of Washington, Seattle, WA 98195-1580, USA
\and
Department of Astronomy, Williams College, Williamstown, MA 01267, USA
\and
INAF-Osservatorio Astrofisico di Arcetri, Largo E. Fermi, 5, I-50125 Firenze, Italy
\and
Centro de Astrobiolog\'ia (CAB, CSIC-INTA), ESAC Campus Camino Bajo del Castillo, s/n, Villanueva de la Ca\~nada, E-28692 Madrid, Spain
\\
\email{amr@iac.es}}

\authorrunning {Galera-Rosillo et al.}
\titlerunning {On the most luminous planetary nebulae of M31}
   \date{\today}

 
  \abstract
   {The planetary nebula luminosity function (PNLF) is a standard candle that comprises a key rung on the extragalactic distance ladder. The method is based on the empirical evidence that the luminosity function of planetary nebulae (PNe) in the [O~{\sc iii}] $\lambda5007$ nebular emission line reaches a maximum value that is approximately invariant with population age, metallicity, or host galaxy type. However, the presence of bright PNe in old stellar populations is not easily explained by single-star evolutionary models.  
   }
   {To gain information about the progenitors of PNe at the tip of the PNLF, we obtained 
   the deepest existing spectra of a sample of PNe in the galaxy M31 to determine their physico-chemical properties and infer the post-asymptotic giant branch (AGB) masses of their central stars (CSs). Precise chemical abundances allow us to confront the theoretical yields for AGB stellar masses and metallicities expected at the bright end of the PNLF. Central star masses of the sampled PNe provide direct information on the controversial origin of the universal cutoff of the PNLF.}
    {Using the OSIRIS instrument at the 10.4m Gran Telescopio Canarias (GTC), optical spectra of nine bright M31 PNe were obtained: four of them at the tip of the PNLF, and the other five some 0.5 magnitudes fainter. A control sample of 
    21 PNe with previous GTC spectra from the literature is also included. We analyze their physical properties and chemical abundances (He, N, O, Ar, Ne, and S), searching for relevant differences between bright PNe and the control samples. The CS masses are estimated with Cloudy modeling using the most recent evolutionary  tracks.}
   {The studied PNe show a remarkable uniformity in all their nebular properties, and the brightest PNe show relatively large electron densities. Stellar characteristics also span a narrow range:  $\mathrm{<L_{*}/L_{\odot}>} = 4\,300\pm 310$, $\mathrm{<T_{eff>}} = 122\,000\pm 10\,600$ K for the CSs of the four brightest PNe, and $\mathrm{<L_{*}/L_{\odot}>} = 3\,300\pm 370$, $\mathrm{<T_{eff>}} = 135\,000\pm 26\,000$~K for those in the control set. This groups all the brightest PNe at the location of maximum temperature in the post-AGB tracks for stars with initial masses $\mathrm{M_i=1.5~M_{\odot}}$.}
    {These figures provide robust observational constraints for the stellar progenitors that produce the PNLF cutoff in a star-forming galaxy such as M31, where a large range of initial masses is in principle available. Inconsistency is found, however, in the computed N/O abundance ratios of five nebulae, which are 1.5 to 3 times larger than predicted by the existing nucleosynthesis models for stars of these masses.}

\maketitle

\section{Introduction}

The [O~{\sc iii}] $\lambda5007$ planetary nebula luminosity function \citep[PNLF;][]{Jacoby1980} is the number density of planetary nebulae (PNe) in a stellar system as a function of their luminosity in that specific emission line. It has been measured in more than 50 galaxies, proving to be an important standard secondary candle on the extragalactic distance ladder out to $\sim$20 Mpc \citep[see][and references therein]{Ciardullo2013IAUS}, and it is planned to be extended out to $\sim$50 Mpc \citep{Chase2021} using novel detection and analysis techniques with the VLT Multi Unit Spectroscopic Explorer (MUSE) and similar instruments \citep{spiggs2020,roth2021}

The method is based on the empirical evidence that the luminosity of a statistical set of PNe in the [O~{\sc iii}] $\lambda5007$ nebular emission line reaches a maximum value
that is invariant with galaxy type and only has a mild dependence on metallicity \citep[e.g.,][]{Dopita1992,Ciardullo2010}. This emission includes up to 80$\%$ of the entire light emitted by a PN (typically $\mathrm{L\sim10^{4}~L_\odot}$), which, for the brightest objects, means $\sim$600~L$_\odot$ emitted in just a single line \citep{Schonberner2007}. 

The bright end of the PNLF in Local Group galaxies was first modeled as an exponential by \cite{Ciardullo1989a} as:
$$\mathrm{N(m)\propto e^{0.307m}[1-e^{3(M^{*}-m)}]},$$
where N is the number of PNe in any given [O~{\sc iii}] $\lambda5007$ magnitude bin,  $\mathrm{M^{*}}$ is the bright cutoff magnitude, and $\mathrm{m}$ is the observed [O~{\sc iii}] $\lambda5007$ magnitude of each PN, computed as $$\mathrm{m=-2.5\cdot log(F_{5007})-13.74 },$$ where $\mathrm{F_{5007}}$ is the absolute flux of the nebula in [O~{\sc iii}] $\lambda5007$ at a distance of 10 pc \citep{Jacoby1989}. 
The current estimated value of the PNLF cutoff is $\mathrm{M^{*} = -4.54\pm0.05}$ \citep{Ciardullo2013IAUS}.\\

Internal tests within galaxy clusters, and comparison with distances from Cepheids and surface brightness fluctuations, yield distance values that consistently agree within 10$\%$ \citep{Ciardullo2010}. Furthermore, as PNe descend from progenitors present in any galactic subsystem, the PNLF method is equally effective for both spiral and elliptical galaxies, providing a crucial link between the Population I and Population II distance scales up to 20 Mpc.
However, despite the PNLF-distance method being well tested empirically, it does not rely on a conclusive theoretical foundation \citep{Ciardullo2013IAUS}:  
 According to standard stellar evolution, the bright cutoff (M*) should not be constant within galaxy types, and the brightest PNe should not be found in old populations.

Theoretical modeling \citep[]{Schonberner2007, Mendez2008} predicted that PNe with central star (CS) masses of $\sim$0.6~M$_\odot$ can attain the observed M* if accompanied by a sufficiently delayed optically thick-to-thin transition of the nebular gas. Adopting the empirical initial-to-final mass relationships (IFMRs) of \cite{kalirai2008}, such a relatively high core mass would imply progenitors with an initial mass of $\sim$2~M$_\odot$. But, according to the evolution and visibility times of a PN, such relatively high-mass stars are too scarce (or completely absent) in old stellar systems such as elliptical galaxies or haloes to produce the number of observed bright PNe.

Although evidence points to a constant cutoff, contradictory conclusions have been reached over the years in trying to explain why M* should be constant \citep{Ciardullo1989a, Jacoby1989, Dopita1992, Mendez1993, Marigo2004}.
It is expected that the cutoff is sensitive to population age, but this fading is not observed. Paradigmatic examples are the galaxies NGC 4697 (Hubble type E6) and NGC 5128 (interacting S0p), both of which contain extremely luminous PNe, with maximum absolute magnitudes of $-5.6$ and $-5.7$,  respectively \citep{Mendez2008a, Peng2004}. The massive CSs needed to generate these PNe, with lifetimes < 700 Ma, are not expected to be found in these old red populations.
However, \cite{Gesicki2018Nat} showed recently that less massive stars, $\mathrm{M_{i}\sim 1.1-2.0~M_{\odot}}$, can indeed produce bright PNe through the accelerated-evolution post-asymptotic giant branch (AGB) star models from \cite{Miller_Bertolami2016}. Their model predicts bright PNe able to explain the PNLF and its bright cutoff at smaller CS masses and therefore larger lifetimes, $\sim1-7$ Ga. 

\cite{Davis2018} tested the \cite{Gesicki2018Nat} results in the M31 bulge, showing that lifetimes of stars with masses able to reproduce the observed PN peak luminosities are about 1.5 Ga, too short to be present: More than ten bright PNe are found, but only $\sim$2 are predicted. Moreover, those authors showed that, when a proper extinction correction is accounted for, there are not only bright PNe where they are not expected, but they are substantially brighter than previously believed. Therefore, the question remains open regarding the role of progenitors with $\mathrm{M_{i}\sim 1-2~M_{\odot}}$ in old systems such as the M31 bulge.

Alternatives to explain the formation of bright PNe and the PNLF invariance have been proposed, such as binary evolution \citep[e.g., blue stragglers;][]{Ciardullo2005}, misidentification with other type of objects, or mass accretion in interacting binary stars \citep[symbiotic stars;][]{Soker2006}. Whether these alternative evolutionary channels are able to provide the observed number of M* PNe is, however, not clear.

M31 is the nearest large spiral; it has a fascinating assembly history and harbors the largest collection of PNe found in any galaxy, including the Milky Way. M31 PNe represent an optimum laboratory for studying the PNLF and its still incomplete theoretical foundations \citep{dsouza2018}, attracting observers' attention in recent years \citep[cf.][]{ Kwitter2012,Balick2013,Corradi2015,Henry2018,Fang2015,Fang2018, Davis2018}. In a pioneering paper, \cite{Merrett2006} cataloged 2615 likely PNe, enabling the determination of the PNLF down to $\sim 3.5$ mag from the cutoff. More recently, \cite{Bhattacharya2021} carried out an imaging survey of M31 over a huge area of 54 sq. deg. that included two disk regions and six inner halo substructures. \cite{Bhattacharya2021} increased the number of PN candidates to 5265 (only a small fraction of them being confirmed with spectroscopy) and extended the PNLF up to $\mathrm{m_{5007}}>26$ mag, enabling the determination of individual PN luminosity functions for each substructure. Given the large metallicity range covered by the substructures, different values for the PNLF cutoff, M*, are expected and indeed measured by \cite{Bhattacharya2021}, albeit with a slope steeper than theoretically predicted in the low metallicity regime \citep{Dopita1992}. 

Given the above uncertainties, it is necessary to gain further insight into the nature of PNe at the PNLF tip in different stellar systems.  As \cite{Davis2018} highlight, this problem can also lead to an incorrect estimation of a galaxy's stellar mass, star formation rate, chemical history, and the initial mass function. Even though extragalactic PNe beyond the Magellanic Clouds are spatially unresolved, precluding knowledge of their morphology and the spatial distribution of physico-chemical properties, large telescopes provide the opportunity to obtain deep integrated spectra of the brightest PNe and to determine nebular and stellar parameters with good accuracy. With this goal, we present in this paper a detailed study of a sample of the brightest PNe in the disk of the galaxy M31. 
In Sect.~\ref{sec:obs} the sample and the observations are presented. Section 3 covers the physical conditions of the observed PNe and Sect. 4 the analysis of their chemical properties. Section 5 presents the derivation of the CS parameters of each object using the results from Cloudy and post-AGB evolutionary models. Finally, we present a discussion and the conclusions derived from the study in Sects. 6 and 7. 

\section{Observations and data reduction} 
\label{sec:obs}
The nine target PNe were selected from \cite{Merrett2006}. They are listed in Table~\ref{table:log}. Their $\mathrm{m_{5007}}$ values range between 20.2 and 20.9. Four of the PNe in our sample (those marked with an asterisk in Table~\ref{table:log}) are located on the bright tip of the PNLF. The five remaining objects, which are all within 0.5 mag of the bright tip and therefore still representative of the very bright PNe that are typically observed in external galaxies, are taken in this study as a control sample together with other objects from the literature.

\begin{table}
        \caption{Observing log of the sample PNe.}             
        \label{table:log}      
        \centering                          
        \begin{tabular}{l c c c }      
                \hline\hline                 
                
                NAME   &    RA       &     Dec       & Exposures$^{\rm a}$   \\
                       &  (J2000)    &   (J2000)     &  (sec)       \\
                
                \hline          
                PN1687*  & 00 43 21.3  & +41 05 28.9 &   4 $\times$ 1622. \\
                PN2068*  & 00 40 49.6  & +40 39 46.5 &   4 $\times$ 1622. \\
                PN2538*  & 00 36 28.8  & +39 35 26.4 &   4 $\times$ 1640. \\
                PN50*    & 00 46 42.9  & +42 08 35.3 &   4 $\times$ 1640. \\
                PN1596   & 00 39 22.6  & +41 06 57.3 &   4 $\times$ 1640. \\
                PN2471   & 00 43 11.2  & +42 20 45.7 &   5 $\times$ 1308. \\
                PN2860   & 00 38 55.0  & +41 06 55.3 &   5 $\times$ 1308. \\
                PN1074   & 00 40 38.1  & +41 16 48.0 &   5 $\times$ 1308. \\
                PN1675   & 00 43 15.2  & +41 04 21.5 &   4 $\times$ 1622. \\
                
                \hline                                   
        \end{tabular}
        \begin{flushleft}
$^{\rm a}$ Number of exposures multiplied by exposure time.
        \end{flushleft}

\end{table} 

\begin{table*}
        \caption{Properties of the observed PNe.}             
        \label{table:properties}      .
        \centering                          
        \begin{tabular}{l c c c  c c c c c c}      
                \hline\hline                 
                
                NAME   &  $\mathrm{m_{5007}}$  & $\xi$     & $\eta$     &  $d_{app}$  &  $d_{app}$  &  $d_{disk}$ &   $v_{rad rot}$  &   $v_{obs}$  &   $v_{diff}$ \\
                       &   (mag)   & (deg)     &   (deg)    &  (deg)      &   (kpc)     &  (kpc)      &   (km$\cdot$s$^{-1}$)   &   (km$\cdot$s$^{-1}$)      &     (km$\cdot$s$^{-1}$)    \\
                 
                \hline          
                PN1687*  &  20.16 &   0.12     &  $-$0.18   &  0.21   &  2.82  &  12.69   &  $-$326    &  $-$362   &   $-$36 \\
                PN2068*  &  20.19 &   $-$0.36  &  $-$0.61   &  0.71   &  9.54  &  10.78   &  $-$514    &  $-$423   &    91 \\
                PN2538*  &  20.25 &   $-$1.21  &  $-$1.67   &  2.06   &  27.7  &  27.95   &  $-$541    &  $-$426   &   114 \\
                PN50*    &  20.33 &   0.74     &   0.88     &  1.15   &  15.5  &  15.67   &   $-$79    &   $-$42   &    37 \\
                PN1596   &  20.71 &   $-$0.63  &  $-$0.15   &  0.65   &  8.74  &  26.69   &  $-$369    &  $-$378   &    $-$9 \\
                PN2471   &  20.69 &   0.08     &   1.08     &  1.08   &  14.5  &  39.33   &  $-$237    &  $-$254   &   $-$17 \\
                PN2860   &  20.84 &   $-$0.72  &  $-$0.15   &  0.74   &  9.95  &  31.07   &  $-$366    &  $-$419   &   $-$53 \\
                PN1074   &  20.88 &   $-$0.40  &   0.01     &  0.40   &  5.38  &  20.43   &  $-$345    &  $-$453   &  $-$108 \\
                PN1675   &  20.67 &   0.10     &  $-$0.20   &  0.22   &  2.96  &  12.49   &  $-$333    &  $-$379   &   $-$46 \\
                
                \hline                                   
        \end{tabular}
        
        \begin{flushleft}
                Assumed parameters for M31: \\ Distance = 770 kpc \citep{Freedman1990}. Center: R.A. (J2000) = 00 42 44.3, Dec. (J2000) = +41 16 09.0 \citep{Merrett2006}. Disk inclination: i = 77$^{\circ}$.7. Position angle = 37$^{\circ}$.7 \citep{Merrett2006}. Heliocentric system velocity: V$_{subsys}$=$-$309 km $\cdot$ s$^{-1}$
        \end{flushleft}
        
\end{table*} 

Spectroscopy of these PNe was secured on different nights in September 2015 in service queue mode at the 10.4m Gran Telescopio Canarias (GTC) telescope on the island of La Palma, Spain. The Optical System for Imaging and low-Intermediate-Resolution Integrated Spectroscopy
(OSIRIS) spectrograph \citep{cepaetal00,cepaetal03} was used in its long-slit mode, along with the R1000B grism and a slit width of 0.8$''$, resulting in spectral coverage from 3630 to 7850 \AA\ with a reciprocal spectral dispersion of 2.1 \AA~pix$^{-1}$ and a spectral resolution $R=6.3$ \AA.
During the observations, the seeing ranged from 0.6$''$ to 0.9$''$, and the data were obtained under photometric conditions and grey moon. 
The total exposure time per target is indicated in Table~\ref{table:log}. 
The GTC calibration plan provided the G191-B2B \citep{Oke1990} spectrophotometric standard star for each observing night.
To minimize the effects of atmospheric differential refraction, the long slit was oriented along the parallactic angle for all the PNe and standard stars.

In Table~\ref{table:properties} the basic parameters adopted for M31 and the observed PNe are listed as follows:
Col. (2), $\mathrm{m_{5007}}$ as determined by \citet{Merrett2006} with quoted typical uncertainty of $\sim$0.07 mag; Cols. (3) and (4), RA ($\xi$) and Dec. ($\eta$) offsets relative to the center of M31; Cols. (5) and (6), total apparent angular distance from the center, d$_{app}$, in deg and kpc, respectively; Col. (7), de-projected distance in the plane of the disk, $d_{disk}$; Col. (8), radial velocities adopted from \cite{Merrett2006} with a typical uncertainty of $\sim$14 km s$^{-1}$; Col. (9), expected radial velocity according to the kinematic model of the extended disk of \cite{Ibata2005}; and Col. (10), differences between the measured (Col. 8) and expected (Col. 9) velocities.

\begin{figure}
        \centering
        {\includegraphics[width=9cm]{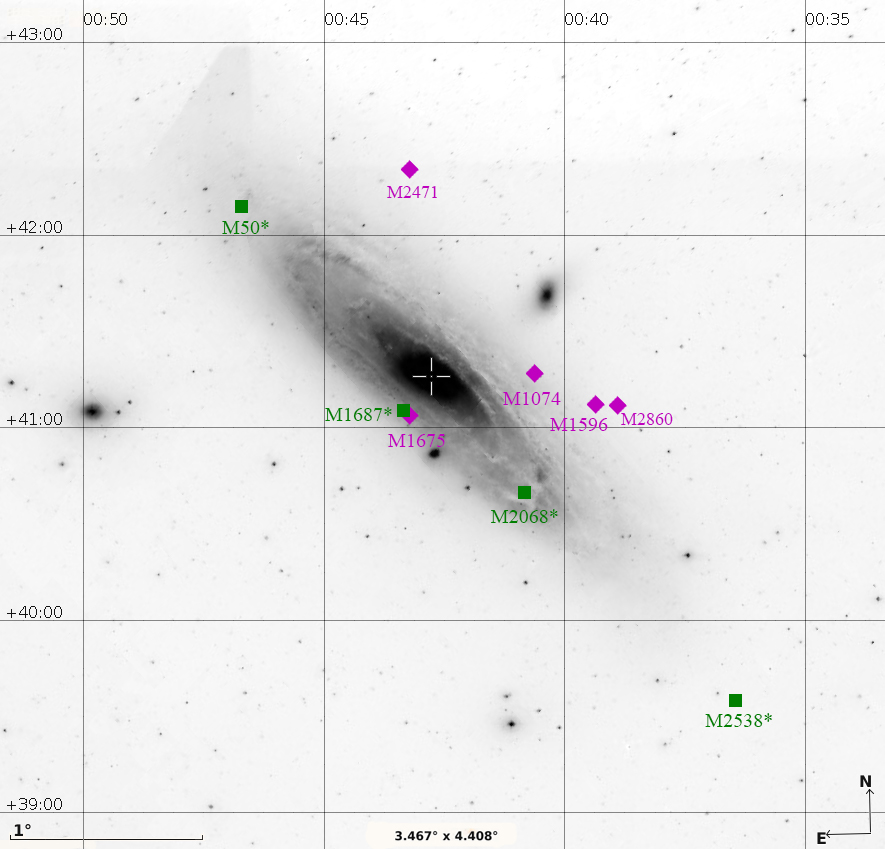}}
        \caption{\label{Fig:location}Location of the target PNe. North is up, and east is to the left. Green squares correspond to the brightest sample targets and magenta diamonds to the control sample. More details are given in the text.}
\end{figure}

Small velocity differences in Col. (10) of Table~\ref{table:properties} give further confidence that the observed PNe belong to the disk of M31, large ones ($\ge$100~km s$^{-1}$ such as for PNe M2538* and M1074) may indicate that these PNe are associated with other M31 substructures, also considering their large de-projected galactocentric distances. Positions on the sky of the target PNe are shown in Fig. \ref{Fig:location}.  

\begin{figure*}
        \centering
        \subfigure[\hbox{GTC OSIRIS 1D spectrum of the bright M50* PN.}]
        {\includegraphics[width=18cm]{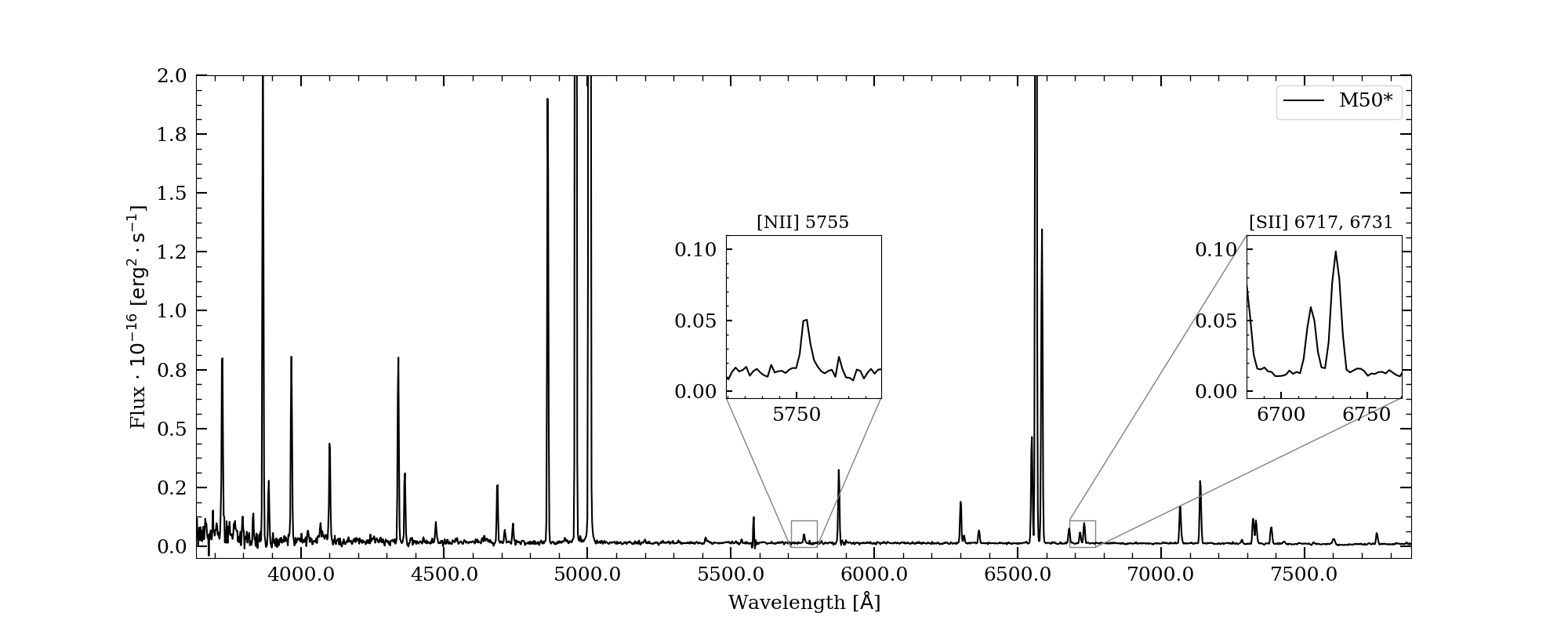}}
	
	\vskip1mm
        \subfigure[\hbox{GTC OSIRIS 1D spectrum of the control M1074 PN.}]
        {\includegraphics[width=18cm]{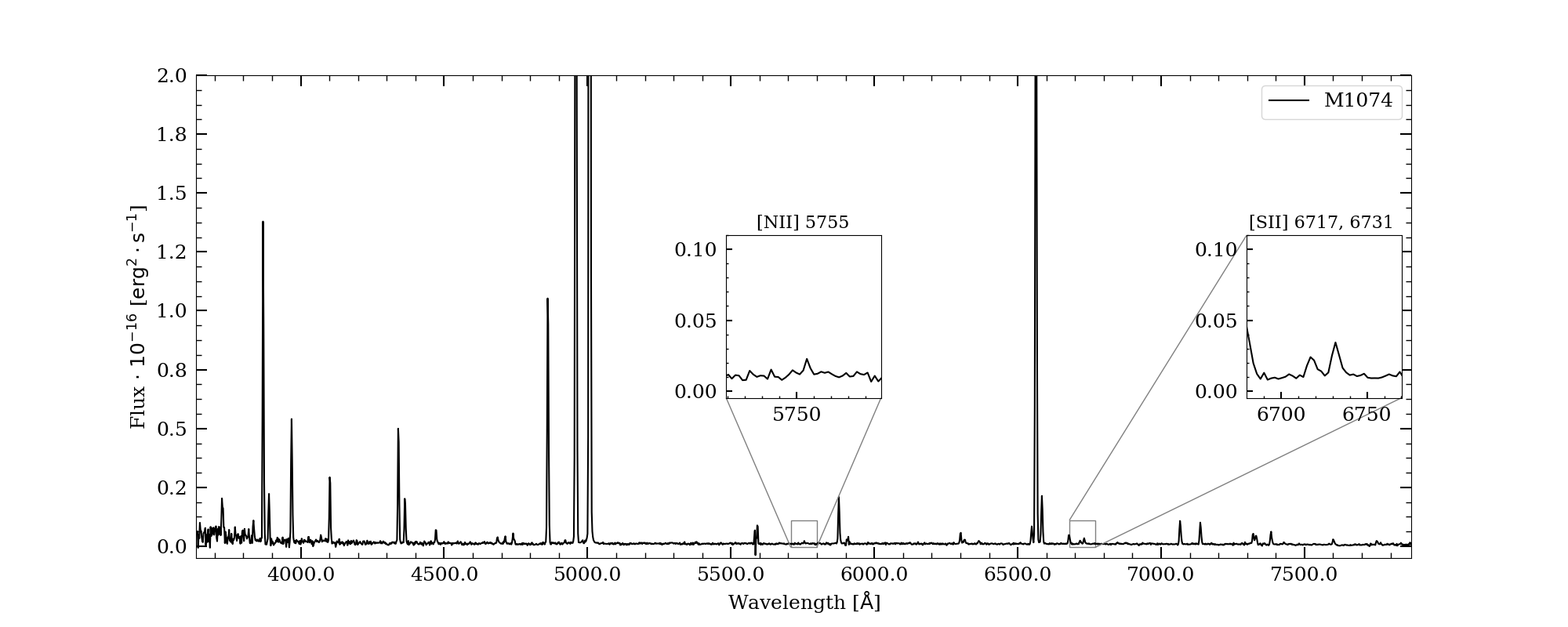}}
        \caption{\label{Fig:spectrum} Spectra of two of the sampled PNe plotted on the same scale and showing the best and worst S/N achieved in key diagnostic faint lines.}
\end{figure*}

The GTC spectroscopic data were reduced following the standard procedure using IRAF\footnote{The Image Reduction and Analysis Facility (IRAF) is distributed by the National Optical Astronomy Observatories, which is operated by the Association of Universities for Research in Astronomy, Inc. (AURA) under cooperative agreement with the National Science Foundation}. Spectra were bias-subtracted, flat-fielded, and wavelength-calibrated in 2D. Cosmic rays were removed using the \textit{lacos} script \citep{lacos2001}, and then the sky background was subtracted. Finally, the spectra were flux calibrated and 1D spectra were extracted using \textit{twodspec} through the traces of the spectrophotometric standards, given that, at the M31 distance, PNe are spatially unresolved.
As an example, two fully calibrated 1D spectra are shown in Fig.~\ref{Fig:spectrum}.

\section{PN physical conditions and chemical abundances} 

\subsection{Emission line fluxes and reddening measurements}
\label{sec:flux_exticnt}

The emission line fluxes and the corresponding statistical errors were measured from the extracted 1D spectra by multi-Gaussian fit using the \textit{splot} task in IRAF. 
The fluxes need to be corrected for interstellar extinction, which is one of the primary sources of uncertainty linked to the PNLF. Extinction comes from different contributions, which cannot always be determined empirically. In general, the interstellar extinction all along the line-of-sight up to each M31 PN CS is composed of both galactic and local contributions due to i) dust in our own Galaxy that may change significantly with the position on M31, ii) dust in M31 itself, all along the path up to the PN, iii) circumnebular dust (mainly from mass ejected during the early AGB phase), iv) intranebular dust, mixed with the ionized gas inside the PN shell, and v) circumstellar dust (i.e., that located around and close to the CS).
By measuring the extinction from the logarithmic extinction parameter, c(H$\beta$), contributions i) to iv) are accounted for, providing the extinction to each PN. 

The c(H$\beta$) value was derived for each PN by comparing the observed and theoretical H$\alpha$/H$\beta$ ratio \citep{Storey_Hummer1995} adopting the intrinsic ratio $I({\rm H}\alpha)/I({\rm H}\beta)$  of 2.86 corresponding to $\mathrm{T_{e}}$ = 10\,000 K and $\mathrm{n_{e}}$ = 1\,000 cm$^{-3}$. We used the extinction curve from \cite{Cardelli1989} with $R_{v}=3.1$. 
The observed flux ratio, extinction coefficient c(H$\beta$), 
de-reddened magnitude ($m_{5007}$), and corresponding absolute magnitude ($M_{5007}$) for each PN in the sample, are presented in Table \ref{table:extinction}. The mean value found, $<$c(H$\beta$)$>$=0.13, and its range, 0.00-0.23 mag, are similar to those of PNe located at the outer disk and external regions of M31 by \citet[][$<$c(H$\beta$)$>$=0.15; range from 0.01 to 0.36]{Kwitter2012}, and \citet[][$<$c(H$\beta$)$>$=0.16; range from 0.06 to 0.29]{Corradi2015}, indicating a dominant interstellar, not local, origin.

In the cases where line fluxes are contaminated by almost coincident He~{\sc i}, He~{\sc ii} or H~{\sc i} lines, and these contaminants account for less than 2$\%$ of the flux, they are considered negligible.
However, in some cases, for example, the [Ar~{\sc iv}] $\lambda$4711 line, blended with the He~{\sc i} 4713 line, or [Ne~{\sc iii}] 3968, blended with He~{\sc i} $\lambda$3965 and H~{\sc i} $\lambda$3970 lines, the flux contribution of the contaminant lines can reach more than  30$\%$ for some of the PNe. The contribution of H~{\sc i} and He~{\sc i} lines to the blends was therefore computed from  theoretical ratios with well isolated H~{\sc i} and He~{\sc i} lines, using the effective recombination coefficients shown in Table~\ref{table:atomic_data_rec} by adopting a first estimation of the electron temperature and density of each PN.

\begin{table}
        \caption{H$\alpha$/H$\beta$ ratio, extinction coefficient, and apparent and absolute magnitudes, both corrected for extinction.}          
        \label{table:extinction}      
        \begin{center}                          
        \begin{tabular}{l| c c c c c}      
                \hline\hline   
                ID   & H$\alpha$/H$\beta$ & c(H$\beta$) &  $m_{5007}$   & $M_{5007}$\\
                \hline
                
                M1687*   & 3.37 &  0.20 $\pm$ 0.06  &   19.60   &  -4.83    \\
                M2068*   & 3.13 &  0.13 $\pm$ 0.06  &   19.83   &  -4.60    \\ 
                M2538*   & 2.80 &  0.00 $\pm$ 0.00  &   20.25   &  -4.18   \\
                M50*     & 3.27 &  0.19 $\pm$ 0.06  &   19.78   &  -4.65   \\                     
                M1596    & 3.21 &  0.17 $\pm$ 0.06  &   20.23   &  -4.20   \\                
                M2471    & 2.95 &  0.04 $\pm$ 0.06  &   20.58   &  -3.85  \\ 
                M2860    & 3.06 &  0.10 $\pm$ 0.06  &   20.56   &  -3.87  \\     
                M1074    & 3.07 &  0.11 $\pm$ 0.06  &   20.57   &  -3.86   \\
                M1675    & 3.35 &  0.23 $\pm$ 0.06  &   20.03   &  -4.40   \\
                \hline
        \end{tabular}
        
        \end{center}
\end{table}

\begin{table}
        \caption{Atomic data used for recombination lines.}             
        \label{table:atomic_data_rec}      
        \centering                          
        \begin{tabular}{l l c}      
                \hline\hline   
                Ion & Effective Recombination Coeff. & Case \\
                \hline 
                
                H$^+$    & \cite{Storey_Hummer1995} &  B \\ 
                He$^+$    & \citeauthor{Porter2012}(\citeyear{Porter2012},
                \citeyear{Porter2013}) &   B \\  
                He$^{2+}$ & \cite{Storey_Hummer1995} &  B \\  
                \hline
                
        \end{tabular}
\end{table} 

Total errors were determined from the quadratic propagation of the measured statistical errors of the spectra, which in general decrease from blue toward red, plus an estimated additional 3$\%$ of the measured flux in order to account for systematic errors, which include continuum determination, flux calibration and, although less important, wavelength calibration uncertainties. The error introduced by the extinction correction is also propagated. Possible errors coming from the decontamination of blended lines are not included.
Measured and de-reddened line fluxes normalized to H$\beta$ = 100 with their corresponding uncertainties are presented in the appendix (Table~\ref{table:fluxes_one_table}). The [O~{\sc iii}] $\mathrm{\lambda5007}$ line is saturated in the four brightest objects. The adopted fluxes included in that Table have been obtained from the theoretical relation between the [O~{\sc iii}] $\lambda5007$ and  $\lambda4959$ lines \citep[2.98, according to][]{Storey2000}.

\subsection{Physical conditions}

Electron temperature and density, $\mathrm{T_{e}}$ and $\mathrm{n_{e}}$, can be determined in all nebulae from collisionally excited lines (CELs) for both high and low-excitation regions of the nebulae. Physical conditions were computed  with {\sc PyNeb} \citep{Luridiana2015}, version 1.1.7. 
Diagnostic errors are computed via Monte Carlo simulations with 1500 random values generated for each line flux using a Gaussian distribution centered in the observed intensity. The selected number of random values was established when the errors converged.

Given the available diagnostic lines, two ionization zones, with different temperatures and densities \cite[see, e.g.,][]{Garciarojas2018} are assumed:
a low-ionization region characterized by an ionization potential IP$\leq17$~eV, and a high-ionization region with IP$>17$~eV.
We consider $\mathrm{T_{e}}$ derived from the  [O~{\sc iii}] $\lambda4363/\lambda4959$ ratio, T$_{e}$([O~{\sc iii}]), representative of the high-ionization zone, while $\mathrm{T_{e}}$ determined using the [N~{\sc ii}] $\lambda5755/\lambda6548$ ratio, T$_{e}$([N~{\sc ii}]), would be representative of the low-ionization region.
In the case of the electron density, $\mathrm{n_{e}}$ obtained from the [S~{\sc ii}] 6716/6731 lines corresponds to the low-ionization zone, while for the high-ionization region the available diagnostic ratio is [Ar~{\sc iv}] 4711/4740. The resulting temperatures and densities are shown in Table~\ref{table:TemDen}. Diagnostic $\mathrm{T_{e}}$ versus $\mathrm{n_{e}}$ diagrams are shown in  Figs.~\ref{Fig:plasma_diag_b} and \ref{Fig:plasma_diag_cs} in the appendix.

While the T$_{e}$([O~{\sc iii}]) determination is accurate in all PNe, 
T$_{e}$([N~{\sc ii}]) has in some cases significant errors.  
An extensive empirical study of the temperature structure in PNe made by \citet{Kaler1986} showed that $\mathrm{T_{e}}$ measured for low-ionization species in an homogeneous nebula were similar or lower than the temperature for the high-ionization zone. However, this is not always obtained in practice, as it occurs in some of our PNe in Table~\ref{table:TemDen}, where T$_{e}$([N~{\sc ii}]) > T$_{e}$([O~{\sc iii}]). Previous works \citep[e.g.,][]{Corradi2015,Fang2018} obviate this issue by adopting a constant T$_{e}$([N~{\sc ii}]) of 10300 K \citep[cf.][]{Kaler1986, magrini2009} when He~{\sc ii} $\mathrm{\lambda 4686}$ line is detected (as is true for all our sample). In a recent review, \cite{Morisset2017IAUS} showed that high density clumps in PNe can lead to a possible overestimation of the [N~{\sc ii}] temperature due to an incorrect treatment of the collisional de-excitation of the upper level where the auroral [N~{\sc ii}] $\lambda$5755 line arises. 
In our work the T$_{e}$([O~{\sc iii}]) has been assumed to be the representative of the electron temperature for the whole nebula if uncertainties in T$_{e}$([N~{\sc ii}]) are higher than $\sim$2000 K (that is the case for M1687* and M1074). In all other objects, both  T$_{e}$([O~{\sc iii}]) and T$_{e}$([N~{\sc ii}]) have been used for the temperatures of the corresponding zones, even if T$_{e}$([N~{\sc ii}]) $>$ T$_{e}$([O~{\sc iii}]). 

\begin{table*}
        \caption{Electron temperatures and densities.}            
        \label{table:TemDen}      
        \centering                          
        \begin{tabular}{l c c c c }      
                \hline\hline   
                Parameter   & Diagnostic ratio                                &  M1687*                      &    M2068*                              &    M2538*             \\
                \hline
                $\mathrm{T_{e}}$ (K) &   [O~{\sc iii}] (4959/4363)            & 12710 $\pm$ 350              & 10620 $\pm$ 260                        & 12120  $\pm$ 260  \\
                        \vspace{1.5mm}
                &   {[N~{\sc ii}]} (5755/6548)                                &  21610 $\pm$ 4600            & 10350 $\pm$ 800                        & 11930  $\pm$ 950      \\
                $\mathrm{n_{e}}$ ($cm^{-3}$) &  [S~{\sc ii}] (6716/6731)      & 4610  $\pm_{2350}^{4800}$    & 19500 ::             & 3350 $\pm_{930}^{1280}$       \\
                &  {[Ar~{\sc iv}]} (4711/4740)                                & 30970 $\pm_{8060}^{10900}$   & 35500 $\pm_{7950}^{12200}$             & 6930 $\pm_{3270}^{2220}$      \\     
                &  {[Cl~{\sc iii}]} (5518/5538)                               & ---                         &      ---                                & ---       \\ 
                
                \hline   
                            &                                                   & M50*                            & M1596                         &  M2471      \\ 
                \hline
                $\mathrm{T_{e}}$ (K) &  [O~{\sc iii}] (4959/4363)               & 11480 $\pm$ 220              &  11930 $\pm$ 320                & 11810 $\pm$ 240         \\ 
                        \vspace{1.5mm}
                &  {[N~{\sc ii}]} (5755/6548)                                   & 12750 $\pm$ 1200              &  11700  $\pm$ 940               & 10930 $\pm$ 890      \\ 
                $\mathrm{n_{e}}$ ($cm^{-3}$) &  [S~{\sc ii}] (6716/6731)        & 6460  $\pm_{2550}^{4210}$     & 3530 $\pm_{1125}^{1650}$        & 3220 $\pm_{1260}^{2070}$     \\ 
                &  {[Ar~{\sc iv}]} (4711/4740)                                  & 13340 $\pm_{3500}^{4740}$     & 6040$\pm_{1390}^{1810}$          & 4780 $\pm_{1420}^{2020}$      \\ 
                &  {[Cl~{\sc iii}]} (5518/5538)                                 &  24540 ::                      &  22010 ::          &  ---         \\ 
                \hline   
                         &                                                    &    M2860                  &       M1074                          &    M1675            \\              
                \hline
                $\mathrm{T_{e}}$ (K)  &  [O~{\sc iii}] (4959/4363)            &  10010 $\pm$ 190          &    11820 $\pm$ 260                   & 10550 $\pm$ 450     \\              
                \vspace{1.5mm}
                &  {[N~{\sc ii}]} (5755/6548)                                 &   12540 $\pm$ 1310        &    12820 $\pm$ 2710                 & 11400 $\pm$ 1150     \\              
                $\mathrm{n_{e}}$ ($cm^{-3}$) &  [S~{\sc ii}] (6716/6731)      & 3420 $\pm_{1300}^{2100}$  &    2770  $\pm_{1390}^{2780}$         & 24300 ::   \\              
                &  {[Ar~{\sc iv}]} (4711/4740)                                & 9170 $\pm_{2190}^{2880}$  &    22410  $\pm_{5130}^{6650}$        & 8900  $\pm_{4000}^{7300}$   \\   
                &  {[Cl~{\sc iii}]} (5518/5538)                               &  9330 ::                   &      ---                            & ---       \\ 
                 \hline         
        \end{tabular}
        \begin{flushleft}
                
                 Note.  A double colon (``::'')indicates large uncertainties.
        \end{flushleft}
\end{table*}

In the case of the density, a two-zone scheme was also adopted as the more realistic situation, and both $\mathrm{n_{e}}$ from [Ar~{\sc iv}] and [S~{\sc ii}] line ratios have been used in all the cases except in PNe M2068 and M1675, where the [S~{\sc ii}] density diagnostic is saturated and hence, $\mathrm{n_{e}}$ from [Ar~{\sc iv}] diagnostic is adopted. 

Although density estimations using [Cl~{\sc iii}] 5518/5538 ratio were also obtained for three objects, we opted for not using them given their large uncertainties (see Table~\ref{table:TemDen}).
We emphasize that only PNe M1074 and M2860 seem to show well-differentiated density zones, while for the remaining PNe, both density diagnostics show similar values within the uncertainties. 

\subsection{Ionic and total abundances}
\label{sec:ionTot}

Ionic and total abundances were computed using {\sc PyNeb} from the fluxes reported in Table~\ref{table:fluxes_one_table} of the appendix and the adopted $\mathrm{T_{e}}$ and $\mathrm{n_{e}}$ in Table~\ref{table:TemDen}. 
We used the combination of atomic data sets shown in Table~\ref{table:atomic_data_cels} for CELs and Table~\ref{table:atomic_data_rec} for recombination lines.
Errors in line fluxes and physical conditions have been propagated via Monte Carlo simulations using 1500 random values. 

\begin{table*}
        \caption{Atomic data used for CELs.}             
        \label{table:atomic_data_cels}      
        \centering                          
        \begin{tabular}{l c c}      
                \hline\hline   
                Ion & Transition probabilities & Collision strengths \\
                \hline 
                
                $\mathrm{N^+}$    & \cite{Fischer2004} & \cite{Tayal2011} \\ 
                $\mathrm{O^+}$    & \cite{Fischer2004} & \cite{Kisielius2009} \\  
                $\mathrm{O^{2+}}$ & \cite{Fischer2004} & \cite{Storey2014} \\  
                              &  \cite{Storey2000} &  \\  
                $\mathrm{Ne^{3+}}$ & \cite{Galavis_Mendoza_Zeipen1997} & \cite{Tayal2011} \\  
                $\mathrm{S^+}$ & \cite{Podobedova2009} & \cite{Tayal_Zatsarinny2010} \\  
                $\mathrm{S^{2+}}$ & \cite{Podobedova2009} & \cite{Tayal_Gupta1999} \\  
                $\mathrm{Ar^{2+}}$ & \cite{Kaufman1986}  & \cite{Galavis1995} \\  
                                   & \cite{Mendoza1983} &  \\  
                $\mathrm{Ar^{3+}}$ & \cite{Mendoza_Zeipen1982a} & \cite{Ramsbottom1997}\\  
                $\mathrm{Ar^{4+}}$ & \cite{Mendoza_Zeipen1982b} & \cite{Galavis1995} \\ 
                                  & \cite{Kaufman1986} &  \\ 
                                  &  \cite{LaJohn1993} &  \\ 
                $\mathrm{Cl^{2+}}$ & \cite{Mendoza1983} & \cite{Butler_Zeipen1989} \\ 
                \hline
                
        \end{tabular}
\end{table*} 

Ionic abundances for different ions of O, He, N, S, Ne, Ar and Cl are listed in Table~\ref{table:ionic_abundances} in the appendix.
According to the two-zone scheme described in the previous section, ionic abundances of O$^+$, N$^+$ and S$^+$ were computed for the low-ionization zone, and abundances of He$^+$, He$^{2+}$, O$^{2+}$, Ne$^{2+}$, Ar$^{2+}$, Ar$^{3+}$, Ar$^{4+}$, S$^{2+}$, and Cl$^{2+}$ were computed adopting the measured physical conditions for the high-ionization zone. 
Table~\ref{table:ionic_abundances} presents the adopted ionic abundances considered as the most reliable measure of the ionic abundances. They avoid auroral lines, as well as lines possibly affected by telluric absorption, He or H contamination, or cases with larger errors.

To compute total elemental abundances, the undetected ionization stages of the different elements are corrected by adopting the usual scheme of the ionization correction factors (ICFs). 
Those proposed by \cite{Delgado_Inglada2014} are used for all the elements with the exception of N, for which the classical relation $\mathrm{N/O=N^{+}/O^{+}}$ is applied \citep{peimbertcostero69, Kingsburgh1994}, following the recommendations of \citet{Delgado_Inglada2015}. 
Given the importance of this particular ICF(N) we made the exercise of computing it using \cite{Delgado_Inglada2014} recipe and found that differences were below 22\% in all the objects except for M2860 and M1675, which are very uncertain since their high O$^{2+}$/O ratio fall outside the validity range for this ICF. Moreover, using Machine Learning techniques (C. Morisset, private communication) we estimated the ICFs using the same set of photoionization models as \cite{Delgado_Inglada2014} and considering as an additional constraint the ratio Ar$^{2+}$/Ar$^{3+}$. In this case, the differences with the classical ICF(N) scheme stay below $\sim$25\% except, again, in M2860 and M1675, where differences are much larger, and are also discrepant with the values derived using \cite{Delgado_Inglada2014} prescription. Unfortunately, at this moment there is no alternative way to determine a reliable ICF(N) for objects with very large O$^{2+}$/O, and we decided to retain the same ICF(N) from the classical relation $\mathrm{N/O=N^{+}/O^{+}}$ for all objects.

The computed ICFs are presented in Table~\ref{table:ICF}, and elemental abundances are shown in Table~\ref{table:total_ab}. We also include, for comparison, present Solar System abundances from \citet{lodders21}.

\begin{table*}
        \caption{ Ionization correction factors.}         
        \label{table:ICF}
        \centering                          
        \begin{tabular}{l c c c c c c c c c  }      
                \hline
            \hline   
                Elem.   &    M1687*  &  M2068*   &    M2538*   &  M50*        &    M1596   &   M2471  &   M2860   &  M1074     &    M1675   \\
                \hline
                        He      &    1.      &    1.     &     1.      &    1.        &  1.         & 1.      &  1.       &  1.       &   1.    \\
                O       &    1.01   &  1.02    &   1.06     &    1.07     &  1.19      & 1.19   &  1.03    &  1.01    &   1.05 \\
                N       &    15.7   &  10.4    &   13.6     &    21.3     &  16.5      & 10.9   &  60.4    &  14.8    &   33.1 \\
                Ne      &    1.05   &  1.04    &   1.06     &    1.07     &  1.19      & 1.19   &  1.03    &  1.06    &   1.06 \\
                S       &    1.76   &  1.55    &   1.74     &    1.98     &  2.00      & 1.75   &  2.48    &  1.73    &   2.18 \\
                Ar      &    1.55   &  1.47    &   1.58     &    1.68     &  1.80      & 1.69   &  1.75    &  1.54    &   1.73 \\
                Cl      &     ---    &   ---     &   ---       &    2.06     &  2.12      & ---     &  2.57    &   ---     &   ---    \\
                \hline
                
        \end{tabular}
\end{table*}

\begin{table*}
\caption{Total elemental abundances.}             
\label{table:total_ab}      
\centering    
\resizebox{\textwidth}{!}{
\begin{tabular}{l  c  c  c  c c c c c c c}      
\hline\hline     
  &                   \multicolumn{10}{c}{12+log(X/H)}  \\
\hline                
 Elem    &      M1687*                 &             2068*          &        M2538*               &         M50*              &      M1596                &               M2471          &               M2860             &             M1074          &           M1675           & Solar     \\
\hline                                                                
\vspace{1.5mm}            
He   &  10.98$\pm$0.02   &  11.04$\pm$0.02  &  11.01$\pm$0.02  & 11.04$\pm$0.02  & 11.07$\pm$0.02  &   11.03$\pm$0.02  &    11.05$\pm$0.02     & 11.00$\pm$0.02   &      11.05$\pm$0.02 &  10.92$\pm$0.02  \\
\vspace{1.5mm}
O    &   8.52$\pm_{0.04}^{0.05}$   &   8.71$\pm_{0.05}^{0.06}$  &   8.50$\pm$0.04  &  8.62$\pm$0.04  &  8.65$\pm_{0.04}^{0.05}$  &    8.63$\pm$0.04  &     8.73$\pm_{0.03}^{0.04}$     &  8.51$\pm$0.04   &       8.70$\pm_{0.06}^{0.07 }$  & 8.73$\pm$0.07   \\
\vspace{1.5mm}
N    &   7.72$\pm_{0.09}^{0.10}$   &   8.31$\pm$0.07  &   7.94$\pm_{0.09}^{0.08 }$  &  8.27$\pm$0.08  &  8.41$\pm$0.09  &    7.99$\pm_{0.09}^{0.08 }$  &     8.40$\pm_{0.12}^{0.10}$     &  7.58$\pm_{0.08}^{0.09}$   &       8.75$\pm$0.12  & 7.85$\pm$0.12   \\
\vspace{1.5mm}
Ne   &   7.89$\pm_{0.04}^{0.05}$   &   8.09$\pm$0.05  &   7.72$\pm$0.04  &  7.91$\pm$0.04  &  7.92$\pm$0.05  &    7.91$\pm$0.04  &     7.98$\pm$0.04     &  7.83$\pm$0.04   &       7.93$\pm$0.08  & 8.15$\pm$0.10   \\
\vspace{1.5mm}
S    &   6.51$\pm_{0.08}^{0.07}$   &   6.79$\pm$0.06  &   6.62$\pm$0.07  &  6.66$\pm_{0.06}^{0.05}$  &  6.97$\pm$0.07  &    6.57$\pm$0.08  &     6.98$\pm_{0.08}^{0.07}$     &  6.63$\pm_{0.06}^{0.05}$   &       7.05$\pm$0.13   & 7.15$\pm$0.03  \\
\vspace{1.5mm}
Ar   &   5.87$\pm$0.04   &   6.23$\pm$0.04  &   5.95$\pm$0.04  &  6.19$\pm$0.04  &  6.37$\pm$0.05  &    6.08$\pm$0.05  &     6.32$\pm$0.04     &  5.86$\pm$0.04   &      6.27$\pm$0.06   & 6.50$\pm$0.10  \\
\vspace{1.5mm}
Cl   &   ---                    &   ---                    &   ---                         &  5.13$\pm$0.13  &  5.24$\pm_{0.14}^{0.12}$     &  ---               &     5.20$\pm_{0.13}^{0.12}$     &  ---                     &       ---         & 5.23$\pm$0.06               \\
\hline
\end{tabular}
}
\end{table*}

\subsection{Comparison with previous results}

One PN from our bright sample (M2538*) was observed by \citet{Corradi2015} using the 3.5m APO-ARC telescope, whereas four PNe in the control sample (M1596, M2471, M2860,
and M1074) were observed with the 3.5m APO-ARC and 8.1m Gemini-N telescopes by \citet{Kwitter2012}. The latter correspond to their PN10, PN6, PN15, and PN16, respectively (we note that M1596 was misidentified with M1583: they interchanged PN9 and PN10 IDs). The quality of our spectra supersedes those obtained in previous references, giving more precise physical conditions and chemical abundances. In general, differences in the de-reddened fluxes for the same object in bright and isolated nebular lines as the [O~{\sc iii}]$\lambda$4959 are no bigger than 3$\%$, but in the case of the faint auroral line [O~{\sc iii}] $\lambda$4363 the differences can amount up to $15-30\%$. The differences in the determined physical conditions mainly affect chemical abundances that use ICFs with a strong dependence on the determined ionization degree; this is the case of, for example, N/H, and this is discussed in more detail in Sect.~\ref{sec:nitrogen}.

\section{Chemical abundances in an extended sample of PNe in M31: Correlations among abundance ratios}

To increase the significance of the comparison between PNe at the PNLF tip with less bright PNe (the control sample), an extended sample of PNe in M31 from previous studies is also included. 
Data for 21 PNe with high-quality GTC spectra are taken from \cite{Balick2013}, \cite{Corradi2015}, \cite{Fang2015} and \cite{Fang2018}\footnote{Spectra from \cite{Kwitter2012} lack the [S~{\sc ii}] doublet in 15 of their 16 M31 PNe (they just observed those lines in one PN, which is in fact one of our control sample PNe, M1596). Therefore, the density of these nebulae could not be empirically measured, and the whole sample from  \cite{Kwitter2012} has been excluded from the abundance analysis.}. 
Extinction, electron density and temperature, and chemical abundances of helium, oxygen, and nitrogen have been recalculated from the published line fluxes, in order to have a complete set of homogeneous chemical abundances of those key elements. Abundances for the rest of the elements have been taken straight from the original papers, when available.
We used for all the objects the \cite{Cardelli1989} extinction law, the density $\mathrm{n_{e}}$([S~{\sc ii}]), and the ICF for nitrogen given by the relation N/O=N$^+$/O$^+$. 
When available, $\mathrm{T_{e}}$[O~{\sc iii}] and $\mathrm{T_{e}}$[N~{\sc ii}] are adopted for the high- and low-ionization species, respectively. Otherwise, $\mathrm{T_{e}}$[O~{\sc iii}] has been used for the whole nebula. 

We note that different subsamples were used when analyzing the elemental abundances.
In the case of duplicated objects, we only used the results from the present work. PN M2507 from \cite{Corradi2015} and PN14 from \cite{Fang2018} are the same object, and we adopted the abundances by \cite{Fang2018} owing to the fact that the extinction law applied by these authors is the same as ours. 
PN6 from \cite{Fang2015} and M31-372  from \cite{Corradi2015} are excluded due to the lack of diagnostic sulfur lines.  Finally, PN5 from \cite{Fang2018} was excluded due to the lack of detection of the $\textrm{[O~{\sc ii}]}$ $\lambda$7320 and $\lambda$7330 lines, which precludes proper determination of the nitrogen ICF.

\subsection{Alpha elements}
\label{sec:alpha_e}

\begin{figure*}
	\centering

	\subfigure[] {\includegraphics[width=6cm]{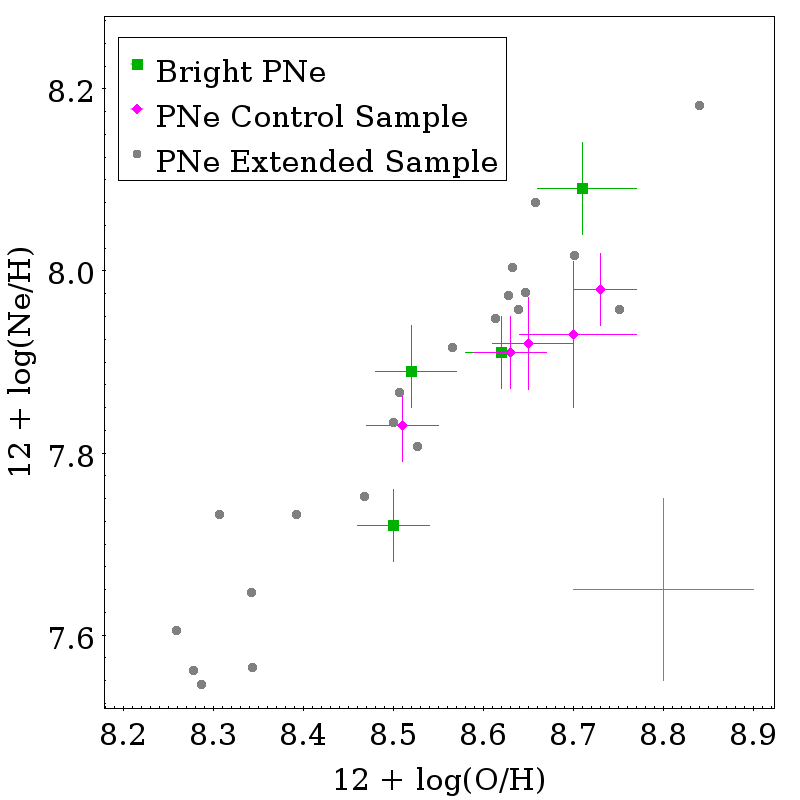}}
	\subfigure[] {\includegraphics[width=6cm]{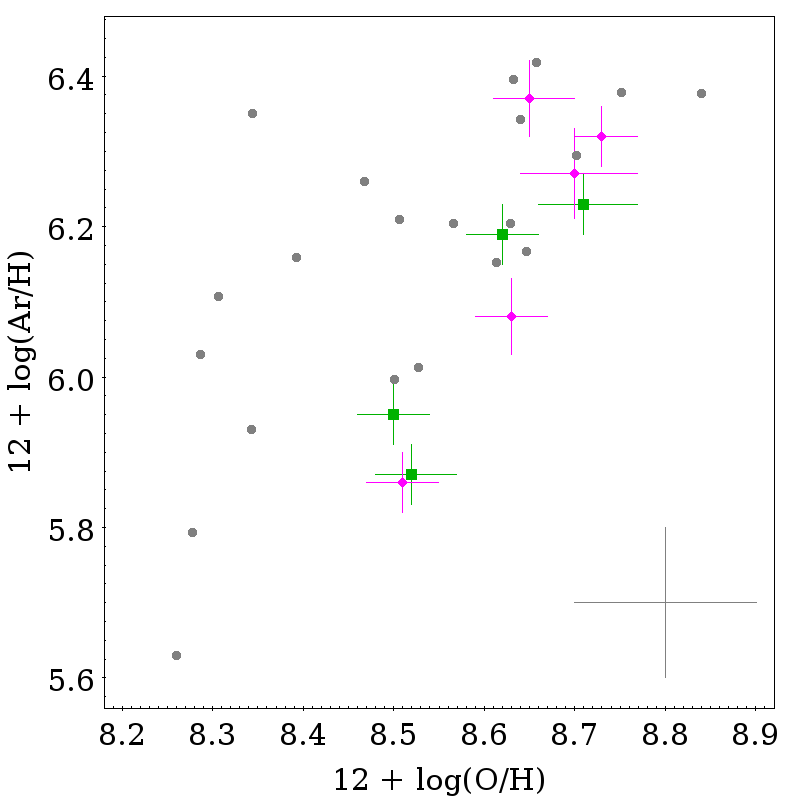}}
	\subfigure[] {\includegraphics[width=6cm]{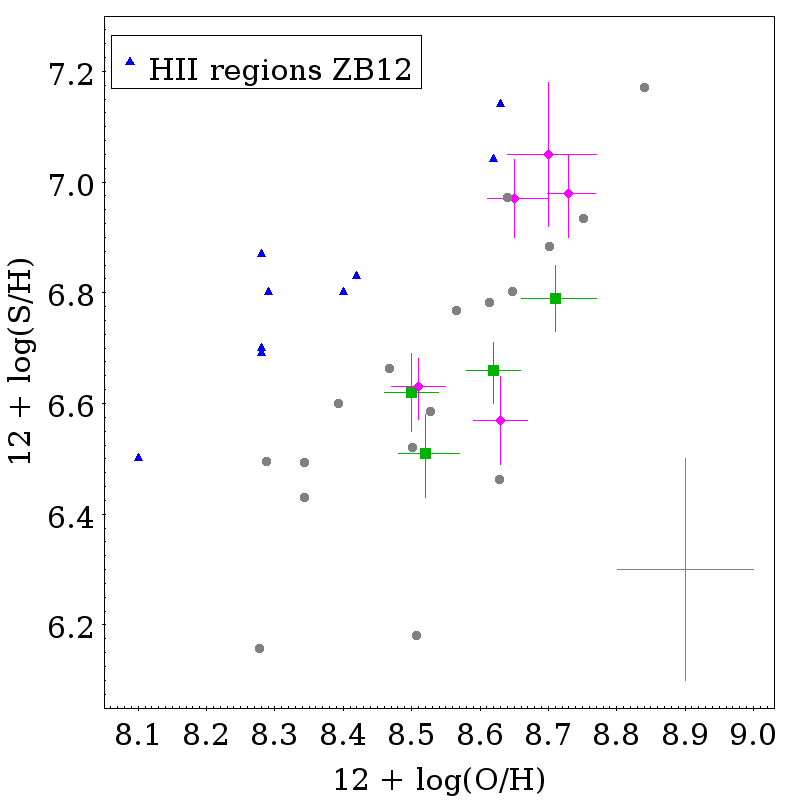}}
	\caption{\label{Fig:Ne_Ar_SvsO}Neon (a), Argon (b), and Sulfur (c) $vs.$ Oxygen abundances for M31 PNe. Green squares correspond to our bright sample and magenta circles to our control sample. Grey circles show the results from the extended sample of M31 PNe from \cite{Balick2013}, \cite{Corradi2015}, \cite{Fang2015}, and \cite{Fang2018}. The grey cross represents the typical uncertainty of the extended sample. Blue triangles in (c) show data for M31 H~{\sc ii} regions from \citet{Zurita2012}.}  
\end{figure*}

All three PN samples show a strong positive correlation of neon, argon and sulfur abundances with oxygen, as shown in Fig. \ref{Fig:Ne_Ar_SvsO} and Table~\ref{table:correlation_coeff}. No significant differences are found among the bright, control, or extended samples, except for a slightly larger dispersion for argon in the extended sample -- but we note that the argon ICF was reported by \citet{Delgado_Inglada2014} as the most uncertain of their ICFs. A tight correlation is expected for these four $\alpha$-elements because they should vary in lockstep after being produced by the same nucleosynthesis processes in massive stars.

The relation between sulfur and oxygen deserves further comment. Figure \ref{Fig:Ne_Ar_SvsO} (c) shows that correlation, including also a sample of M31 H~{\sc ii} regions from \cite{Zurita2012}. 

As with Ne and Ar, neither S nor O are expected to be significantly produced or destroyed during the evolution of low-mass PN progenitors, at least not at the high metallicities of the M31 disk \citep[cf.][]{shingleskarakas13}, but the available data compiled along the years have shown inconclusive results\footnote{\cite{Delgado_Inglada2015} and \cite{garciahernandez2016} report moderate O production, increasing up to 0.3 dex, in carbon-rich, low metallicity Galactic PNe.}. Figure\ \ref{Fig:Ne_Ar_SvsO} (c) shows clearly for the first time the predicted strong  S versus O abundance  correlation for any PN sample (correlation coefficient $r=0.8$; Table~\ref{table:correlation_coeff}). Comparison can be made, for example, with Fig. 8 from \cite{shaw2010}, Fig. 1 from \cite{Henryetal2012}, Fig. 6 from \cite{Maciel2017}, or Fig. 10 from \cite{Fang2018}, where a rough tendency of increasing S versus increasing O abundances for the Milky Way and M31 PNe is apparent, but not quantified. This result gives us confidence on the quality of the observations and the analysis done. On the other hand, the whole sample of PNe is located in Fig.\ \ref{Fig:Ne_Ar_SvsO} (c) just below the area occupied by the H~{\sc ii} regions of \cite{Zurita2012}. We recover here the so-called sulfur anomaly, that is, sulfur abundances in PNe are systematically lower, for a given oxygen abundance, than those found in other probes of interstellar abundances. \citet{Henryetal2012} suggested that this ``anomaly'' could be caused by an incorrect computation of the contribution of ionization stages higher than S$^{2+}$ by commonly used ICF schemes. However, new ICF schemes making use of a large database of photoionization models cannot solve this problem \citep{Delgado_Inglada2014, Delgado_Inglada2017}. 

\begin{figure*}
	\centering

	\subfigure[] {\includegraphics[width=6cm]{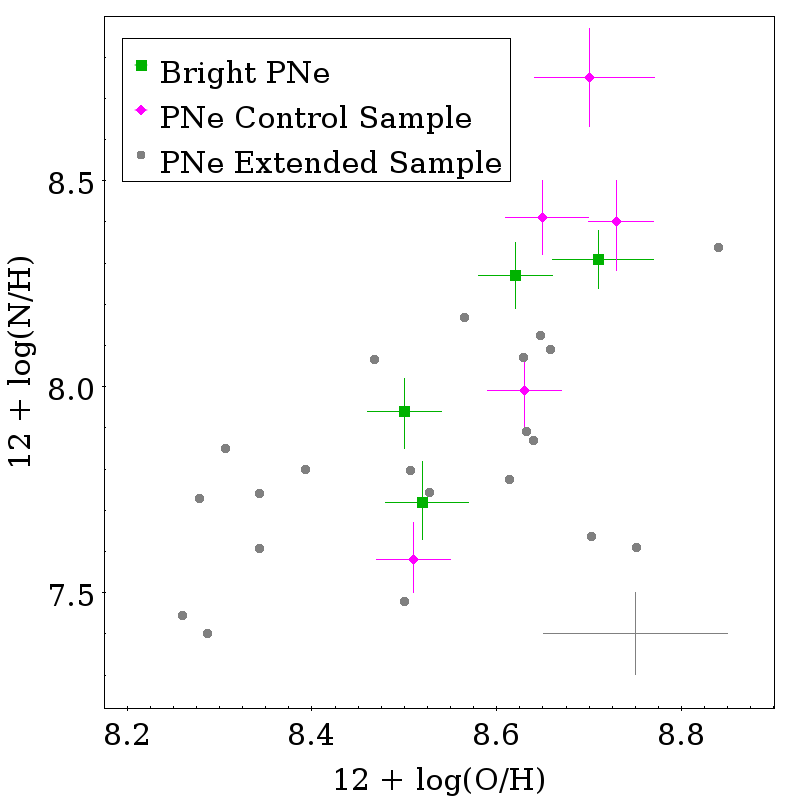}}
	\subfigure[] {\includegraphics[width=6cm]{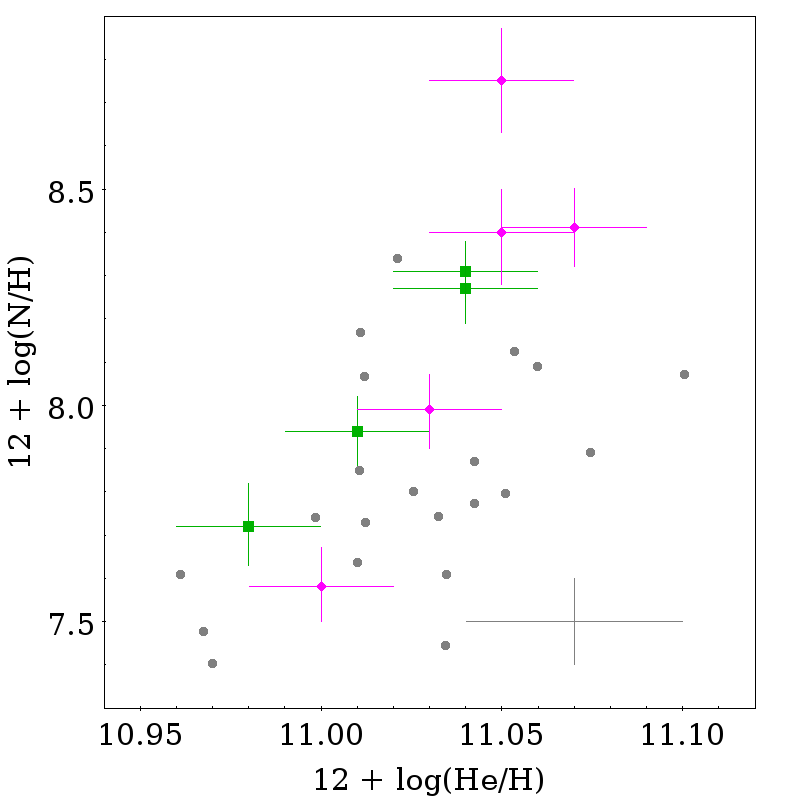}}
	\subfigure[] {\includegraphics[width=6cm]{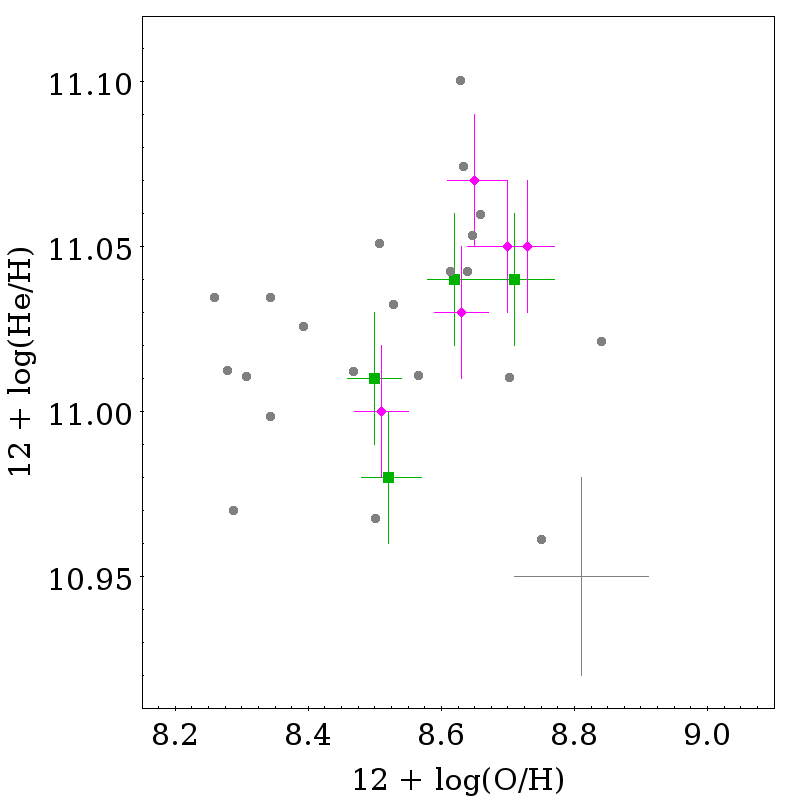}}
		\caption{\label{Fig:N_He_O}Nitrogen {\it vs.} Oxygen (a), Nitrogen {\it vs.} Helium (b), and Helium {\it vs.} Oxygen (c). Labels as in Fig.~\ref{Fig:Ne_Ar_SvsO}.} 
\end{figure*} 

\begin{figure*}
	\centering
	
	\subfigure[] {\includegraphics[width=6cm]{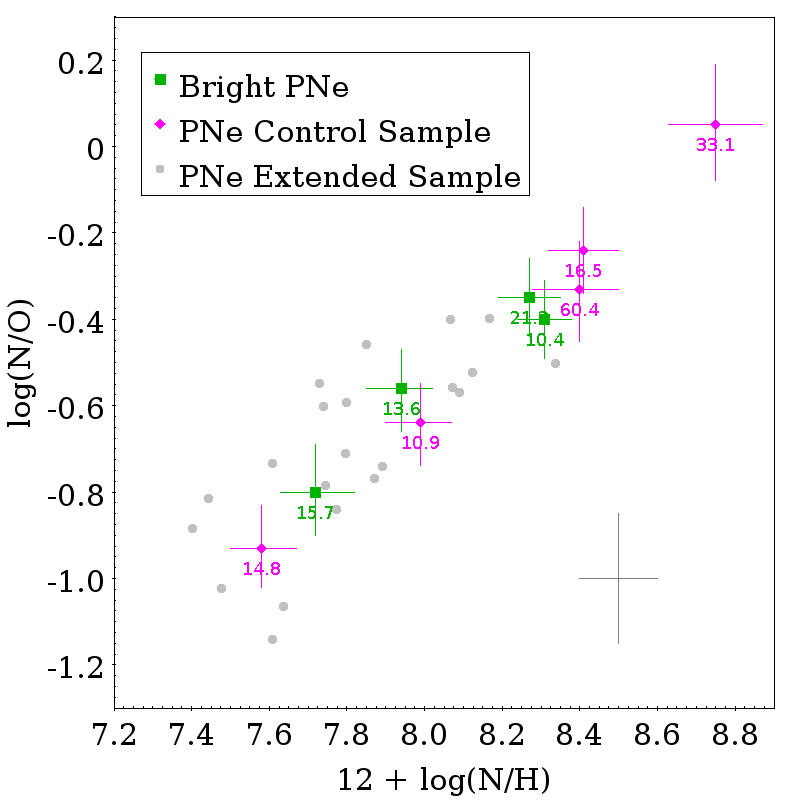}}
	\subfigure[] {\includegraphics[width=6cm]{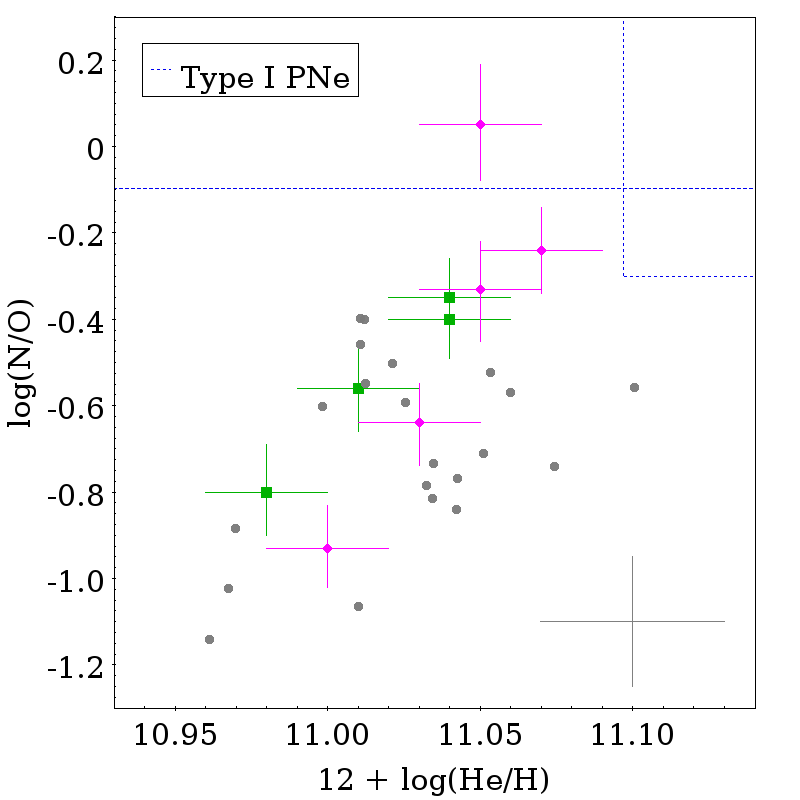}}
	\subfigure[] {\includegraphics[width=6cm]{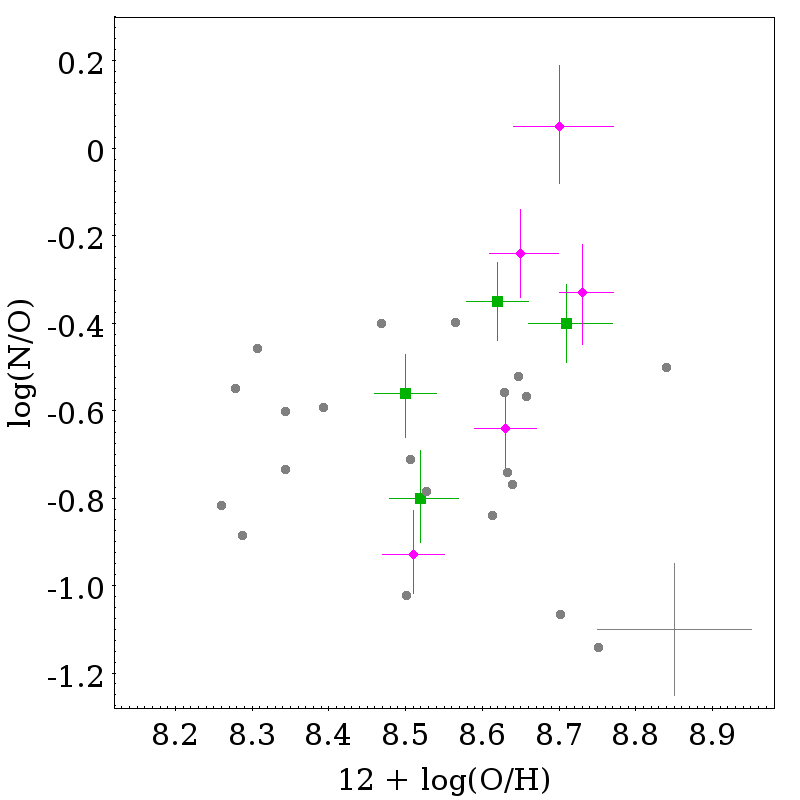}}
	\caption{\label{Fig:NOvsN_He_O}log(N/O) $vs.$ Nitrogen (a), Helium (b), and Oxygen (c) abundances. Labels as in Fig.~\ref{Fig:Ne_Ar_SvsO}. Nitrogen ICF values are labelled in (a) (see text). The dashed rectangle in (b) defines the Type I PNe region from \cite{Peimbert1983}, whereas the horizontal dashed line sets the Type I PNe limit from \citet{Kingsburgh1994}.}
\end{figure*} 

\begin{table*}
\caption{Pearson correlation coefficient, $r$ (and corresponding p values at a significance level of 0.05), for data in Figs. \ref{Fig:Ne_Ar_SvsO}, \ref{Fig:N_He_O}, and \ref{Fig:NOvsN_He_O}.}             
\label{table:correlation_coeff}      
\centering
\small
\begin{adjustbox}{width=\textwidth}
\begin{tabular}{|l|r|r|r|r|r|r|r|r|r|r|}
\hline
  \multicolumn{1}{|c|}{PN Sample} &
  \multicolumn{1}{c|}{n} &
  \multicolumn{1}{c|}{Ne vs O} &
  \multicolumn{1}{c|}{Ar vs O} &
  \multicolumn{1}{c|}{S vs O} &
  \multicolumn{1}{c|}{N vs O} &
  \multicolumn{1}{c|}{N vs He} &
  \multicolumn{1}{c|}{He vs O} &
  \multicolumn{1}{c|}{N/O vs N} &
  \multicolumn{1}{c|}{N/O vs He} &
  \multicolumn{1}{c|}{N/O vs O} \\
\hline
  Bright & 9 & 0.82 & 0.90 & 0.77 & 0.85 & 0.87 & 0.82 & 0.99 & 0.83 & 0.76\\
  +Control & & (0.007) & ($<$0.001) & (0.004) & (0.002) & (0.002) & (0.007) & ($<$0.001) & (0.005) & (0.017)\\
\hline
  Extended & 21 & 0.95 & 0.72 & 0.79$^a$ & 0.55 & 0.49 & 0.23 & 0.74 & 0.39 & $-$0.15\\
  & & ($<$0.001) & ($<$0.001) & ($<$0.001) & (0.011) & (0.024) & (0.32$^b$) & ($<$0.001) & (0.08$^b$) & (0.52$^b$)\\
\hline\end{tabular}
\end{adjustbox}
\begin{description}
                \item $^a$ Only 17 PNe have adequate sulfur data. \\
                \item $^b$ Not significant at level $<$0.05. \\
\end{description}
\end{table*}

\subsection{Nitrogen, helium, and oxygen}
\label{sec:NOHe}

Nitrogen and helium, as well as carbon and s-elements, are substantially synthesized during the evolution of PN progenitors, whereas oxygen, as mentioned earlier, is not. Comparison of N, He, and O relative abundances has been carried out along the years to test nucleosynthesis models and study the progenitors' evolution. Nitrogen, in particular, is a key element in that analysis, but its complex nucleosynthesis makes interpretation intricate.

The relations among N, He, and O abundances for the bright, control, and extended samples are plotted in Fig. \ref{Fig:N_He_O}. A remarkable result is that, while the nine PNe of the bright and control samples are strongly correlated in the three diagrams, (a), (b), and (c), the extended sample of 21 PNe from literature is not. Bright and control PNe appear intertwined in the diagrams and both samples have virtually the same fitting slopes and dispersion.

A mild correlation between N and O was reported early by Pottasch (1983) for Galactic PNe, but only for low (12+log(N/H) $<7.8$) N abundances. Henry (1990) also finds a positive correlation ($r=0.61$) for his non-Type I Galactic PN sample. \cite{Kingsburgh1994}, however, did not find any correlation, and conclude that no significant production of O via the ON cycle occurs during the second dredge-up, even for Type I PNe. More recently, \cite{Maciel2017} claimed a positive correlation between N and O for a large heterogeneous compilation of PNe in the Milky Way and other galaxies, although the correlation is indeed weak (their Fig. 12) and not quantified.
On the theoretical side, it is expected that, for low-intermediate mass progenitors with near solar abundances, the secondary production of N comes at the expenses of C via the carbon-nitrogen-oxygen cycle. An increase in N abundance with metallicity is therefore expected \citep[cf.][] {Magrini2018}. Figure \ref{Fig:N_He_O} (a) shows a strong correlation for our PN sample but a mild one for the extended sample, the latter with a Pearson $r$ coefficient similar to that found by \cite{Henry1990}.

Identical behavior is shown for He and N in Fig. \ref{Fig:N_He_O} (b): a strong correlation for the bright and main PN control samples, but a much weaker one, if any, for the extended sample (Table~\ref{table:correlation_coeff}). As both N and He are produced by the same nucleosynthesis processes, the observed correlation in our data is expected (while the lack of it for the extended sample is striking).

Data in Fig. \ref{Fig:N_He_O} (c) are even more extreme, showing disparate correlation coefficients $r=$0.8 and 0.2 for our nine-PN set and for the extended sample PNe, respectively (Table~\ref{table:correlation_coeff}). Neither Henry (1990) nor \cite{Kingsburgh1994} found any correlation of He versus O in their Galactic PN samples.  In the most recent analysis, 
\citet{Henry2018} compare the observed He and O abundances with the MONASH \citep{Karakas2016} and LPCODE \citep{Miller_Bertolami2016} models, which indeed predict a modest enhancement of He with metallicity.
\cite{Henry2018} argue that observational uncertainties of He/H likely obscure this theoretically predicted trend. Our 12+log(He/H) error bars are typically $\pm0.02$ dex (Table~\ref{table:total_ab}), or $\pm0.005$ in He/H, just at the level suggested by \cite{Henry2018} to allow meaningful comparison of models with observational data. Although the range in He and metallicity of our PNe in M31 is certainly limited, and the slope for the fitting line in Fig. \ref{Fig:N_He_O} (c) is somewhat larger than model results, they do show for the first time the positive He versus O correlation predicted by the models.

\subsection{The nitrogen-to-oxygen ratio}
\label{sec:nitrogen}

The N/O ratio has been extensively used to confront models with observations since it minimizes (together with the C/O ratio) possible effects due to different initial composition of the progenitor stars \citep[cf.][]{Henry2018, Garciarojas2018}. Furthermore, N and O are synthesized by progenitors with very different mass ranges that evolve with different enrichment timescales  \citep[see][]{vincenzo2016, Esteban2020}. For this reason, interpreting N/O data has always been a challenge.

Data for log(N/O) versus 12+log(N/H), 12+log(He/H), and 12+log(O/H) relations are plotted in Figs. \ref{Fig:NOvsN_He_O}\,(a), (b), and (c), respectively. We emphasize that having adopted the ICF(N) from the relation $\mathrm{N/O=N^{+}/O^{+}}$, the derived N/O ratios for all PNe are insensitive to the particular ICF(N) values (Table~\ref{table:ICF}). This is shown in Fig. \ref{Fig:NOvsN_He_O} (a) where ICF(N) values are labeled for the bright and control sets. As expected, a tight correlation between log(N/O) and 12+log(N/H) is observed in Fig. \ref{Fig:NOvsN_He_O}\,(a) for all PN samples, indicating that the variation in N/O is mainly due to variations in nitrogen.

In Fig. \ref{Fig:NOvsN_He_O}\,(b), which displays log(N/O) versus 12+log(He/H), we note that there is no object located in the region of type I PNe following the original definition by \cite{Peimbert1983} (He/H $>0.125$ and log(N/O)$> -0.3$). One of the control sample PNe, M1675, is located at the N/O$ \ge$ 0.8 modified Type I zone defined by \citet{Kingsburgh1994}, but its relatively low He/H abundance casts doubts about being a true Type I nebula. Figure \ref{Fig:NOvsN_He_O}\,(b) provides a first indication about the masses of the progenitor stars: The brightest PNe in M31 are not Type I objects, that is, according to current single-star nucleosynthesis models, they do not come from massive 
($\mathrm{M_i}\ga$2.5~M$_\odot$) progenitor stars \citep [cf.][]{Phillips2001}. 

A strong positive correlation for the bright and control samples ($r=0.8$)
is apparent in Fig. \ref{Fig:NOvsN_He_O}\,(b). Such a correlation is expected since N and O are synthesized independently, while N and He are enhanced during the evolution of all progenitors by the same nucleosynthesis processes. Rough correlations between N/O and He in samples of PNe from the Galaxy, M31 and the Magellanic Clouds were also found by \citet{Kaler1979}, \cite{Jacoby1999}, \cite{shaw2010}, \cite{Maciel2017}, and \cite{Henry2018} (but not by \citealp{Kingsburgh1994}). 

Figure \ref{Fig:NOvsN_He_O}\,(c) shows a strong positive correlation ($r\sim 0.8$) between N/O and O for the bright and control samples.\ However, as in the previous case, it vanishes in the extended sample.

In the case of H~{\sc ii} regions and H~{\sc ii} galaxies, the primary/secondary
production of N is reflected in the observed behavior of N/O versus O: A plateau is generally found at low metallicity followed by a linear rise beyond certain metallicity \citep{Henry2000, Magrini2018}. However, in the Milky Way and M31 only a scattered plateau of H~{\sc ii} regions is apparent up to the highest measured metallicity \citep{Esteban2020, ArellanoCordova2021}. 
For PNe, the situation is even less clear: a poor, if any, correlation between N/O and O has been reported. Data from \cite{Henry1990}, \cite{Kingsburgh1994}, and \cite{Henry2018} show a large dispersion with, at best, a rough increase \citep[or decrease: cf.][]{Jacoby1999, Maciel2009, Milingo2010} of N/O with increasing O abundance.

In PNe with solar-like metallicity and low-mass progenitors, the correlation between N/O and O abundance seen in Fig. \ref{Fig:NOvsN_He_O}\,(c) for the bright and control samples is expected since N is mostly synthesized at the expense of the O and C already present in the star. Furthermore, a slope of N/O versus O steeper for PNe than for H~{\sc ii} regions is also expected since, in a simple closed-box model, the latter depends only on the secular N and O enrichment of the interstellar medium up to the present time. In contrast, the slope for PNe depends on i) the secular N and O enrichment only up to the epoch of the formation of the progenitors (the O abundance, in the abscissa), plus ii) the extra production of N during the evolution of the progenitors (increasing the N/O ratio in the ordinate). Our data indicate a slope around 70{$^\circ$} for the bright and control PN samples, larger than typical values of around 45{$^\circ$} from extragalactic H~{\sc ii} regions at similar O/H abundances \citep{Henry2018}.

\begin{figure}
    {
        \includegraphics[width=9cm]{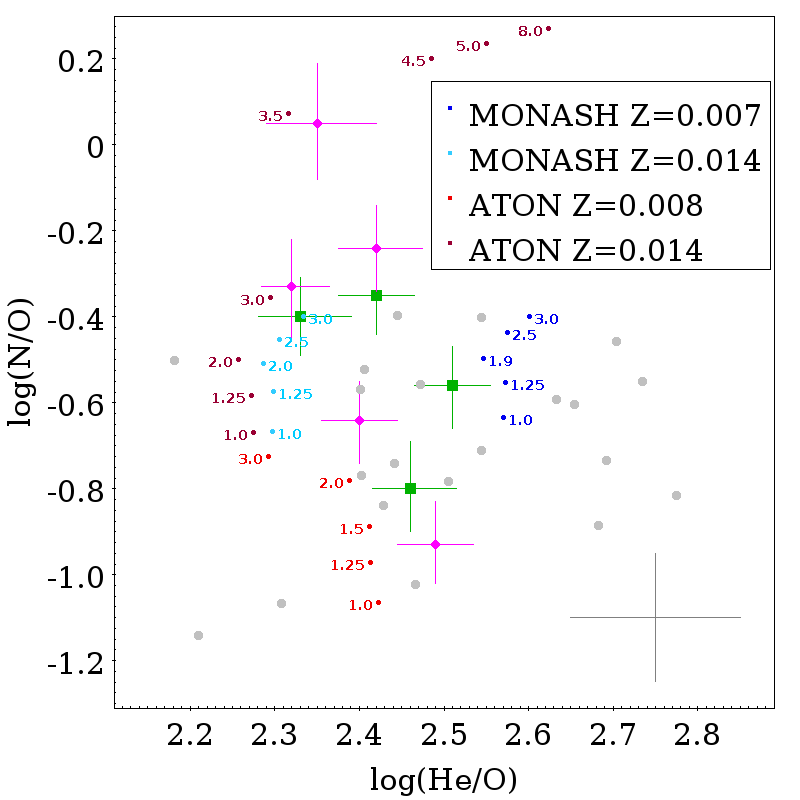}}
        \caption{\label{Fig:NOvsHeO}log(N/O) {\it{vs.}} log(He/O) for the M31 PN samples together with MONASH \citep{Karakas2016} and ATON \citep{Ventura2018} models, labeled with the initial mass of the progenitor. Symbols are the same as in Fig. \ref{Fig:Ne_Ar_SvsO}.} 
\end{figure}

Finally, we compare in Fig. \ref{Fig:NOvsHeO} the observed He/O and N/O ratios with model results from \cite{Karakas2016} (MONASH) and \cite{Ventura2018} (ATON). Metallicities suitable for M31 (i.e., embracing the solar value Z=0.014; \citealt{Saglia2018}) have been chosen.
Once more, the PNe of the bright and control samples appear arranged in a clear anticorrelation, $r=-0.65$, while the extended control sample does not. MONASH results for Z=0.007 and Z=0.014 adequately encompass the observed values of He/O for our nine bright PNe and most of the extended sample, although seven PNe from the latter lie in the zone of lower metallicity, with log(He/O)$>$2.6. None of those PNe are He overabundant Type-I objects, so their positions reflect their lowest O abundance, which is consistent with the fact that five out of those seven outliers are located in areas belonging to the M31 diffuse halo and outer structures. More interesting is the behavior of the N/O ratio. As will be discussed in the next section, the estimated progenitor masses for our nine PNe are restricted to a very narrow range, $1.1\to1.8 ~\mathrm{M_{\odot}}$, and neither MONASH nor ATON models reach the high values of log(N/O)$\ga{-0.4}$ observed in five of our PNe for that range of masses. A similar discrepancy has been reported by, among others, \cite{Henry2018} and points to either a serious limitation of the models (with rotation, magnetic field or extra-mixing not adequately taken into account) or to a wrong determination of the initial/final masses via evolutionary models and the Hertzsprung-Russell (HR) diagram.

In summary, a key characteristic of our data in Figs. \ref{Fig:N_He_O}\,(a), (b), (c), \ref{Fig:NOvsN_He_O}\,(b), (c), and \ref{Fig:NOvsHeO} is the higher degree of correlation of various elemental abundances for the bright and  control samples  with respect to the extended sample. Apart from the smaller uncertainties, our bright and control samples represent an homogeneous population of disk PNe of M31, whereas the extended sample includes a broader range of populations, with distances ranging from $d_{disk}$ $\sim$ 5 to more than 100 kpc and belonging to the halo or to its complex outer substructures, which
span a wide range in metallicity covering roughly one order of magnitude \citep{Bhattacharya2021}.
The location of PNe in these diagrams, and even the PNLF of each population \citep{Bhattacharya2021}, is sensitive to the initial chemical composition, star formation history, and subsequent interaction of each substructure with M31, and that would explain the observed dispersion of the extended sample in the figures.

\section{PN central stars and their progenitors}

\subsection{Cloudy models}

The effective temperature, $\mathrm{T_{eff}}$, and luminosity, $\mathrm{L_{*}}$, of the PN central star (CSPN) locate it in the HR diagram and measure, through the available post-AGB tracks, the final mass $\mathrm{M_{f}}$ of the progenitor star. Then, an initial mass $\mathrm{M_{i}}$ can be derived using available semiempirical IFMRs.

A grid of models was run for each PN using the pyCloudy package \citep{Morisset2013}  based  on  the  1D  photoionization code Cloudy 17.01 version \citep{Ferland2017}.
As starting values for $\mathrm{T_{eff}}$ and $\mathrm{L_{*}}$, we adopted the values estimated with the method outlined in \cite{Dopita1991}, using the PN excitation class (EC) and its $\rm H\beta$ absolute flux. The latter was derived by multiplying the de-reddened spectroscopic  $I(\rm H\beta)$/$I(\lambda5007)$ flux ratios in Table~\ref{table:fluxes_one_table} by the $I(\lambda5007)_{M}$ absolute fluxes from \cite{Merrett2006}. For the four brightest PNe, whose $\lambda5007$ line is saturated, $I(\lambda5007)$ was calculated to be $2.98 \times I(\lambda4959)$ \citep{Storey2000}.  
Excitation class values were obtained for the nine bright and control sample PNe from the $I(\lambda4686)/{I(\rm H}\beta)$ line ratio \citep{Dopita1991}. They show a remarkably narrow range of intermediate ECs, between 4.5 and 6.3. 

The aim of the Cloudy modeling is to match a selection of observed nebular line ratios with those predicted by the model through a given ionization source and a number of assumptions necessitated by the lack of information \citep[see, e.g.,][]{Jacoby1999,Magrini2004,Kwitter2012}.
Cloudy also requires as an input the energy distribution of the CSPN. We used the latest Rauch H-Ni atmospheres with log g = 6.5 or, alternatively, Rauch H-Ca when an effective temperature higher than 190\,000 K is required \citep{Rauch2003}.
Chemical abundances in Table~\ref{table:total_ab} are adopted, and all PNe have been assumed to be optically thick. 
The nebular geometry is assumed to be spherical with an inner radius of $10^{17}$ cm, and the density is set as constant. Even so, the effect of different density distributions has been explored, in particular for cases such as M1687* or M1074 where  different values for the density when using low- or high-excitation ions have been obtained.

The model selection procedure is based mainly on a positive correlation of the He~{\sc ii} 4686 and He~{\sc i} 5876 fluxes with the stellar effective temperature, $\mathrm{T_{eff}}$, whereas the absolute fluxes of ${\rm H}\beta$ and [O{~\sc iii}] $\lambda$5007 constrain the stellar luminosity, $\mathrm{L_{*}}$. Additional line ratios, similar to those investigated in \cite{Kwitter2012}, are also taken into consideration, and further nebular parameters, such as output radii and nebular masses, were always checked to be kept physically reasonable. 

We note that the abundance of carbon, a major nebular coolant, is actually unknown for the PNe in our sample, and this may represent an important source of uncertainty in the modeling. For this reason, values for the carbon abundance corresponding to C/O ratios of 0.5, 1.0, 1.5, and 2.0 have been tested. However, changes in C abundance result in only small variations in the He~{\sc ii} 4686, He~{\sc i} 5876, [O~{\sc iii}] 5007, and ${\rm H}\beta$ diagnostic fluxes, and we finally set a C/O ratio equal to 1.0. In short, although there is no information on the C abundance for our objects, the main emission-line strengths predicted by the models are in good agreement with those observed, and we can be confident of the robustness of the procedure and the resulting stellar parameters.

The $\mathrm{T_{eff}}$ and $\mathrm{L_{*}}$ from Cloudy modeling are listed in Table~\ref{table:m_E_T_L}.
The most remarkable result is the very narrow ranges exhibited by both stellar parameters: log$\mathrm{(L_{*}/L_{\odot})} = 3.57\pm 0.07$, log$\mathrm{T_{eff}} = 5.11\pm 0.07$. All nine CSPNe seem to be caught in a very similar post-AGB evolutionary stage and be produced by similar progenitor stars. Also, as expected, the luminosities $\mathrm{L_{*}}$ of the bright PNe are larger ($\sim 30\%$ on average) than those for the control sample, whereas temperatures $\mathrm{T_{eff}}$ do not show any obvious difference.

\subsection{Location in the HR diagram, progenitor masses, and post-AGB model predictions}

\begin{figure}
        \centering
        \includegraphics[width=9cm]{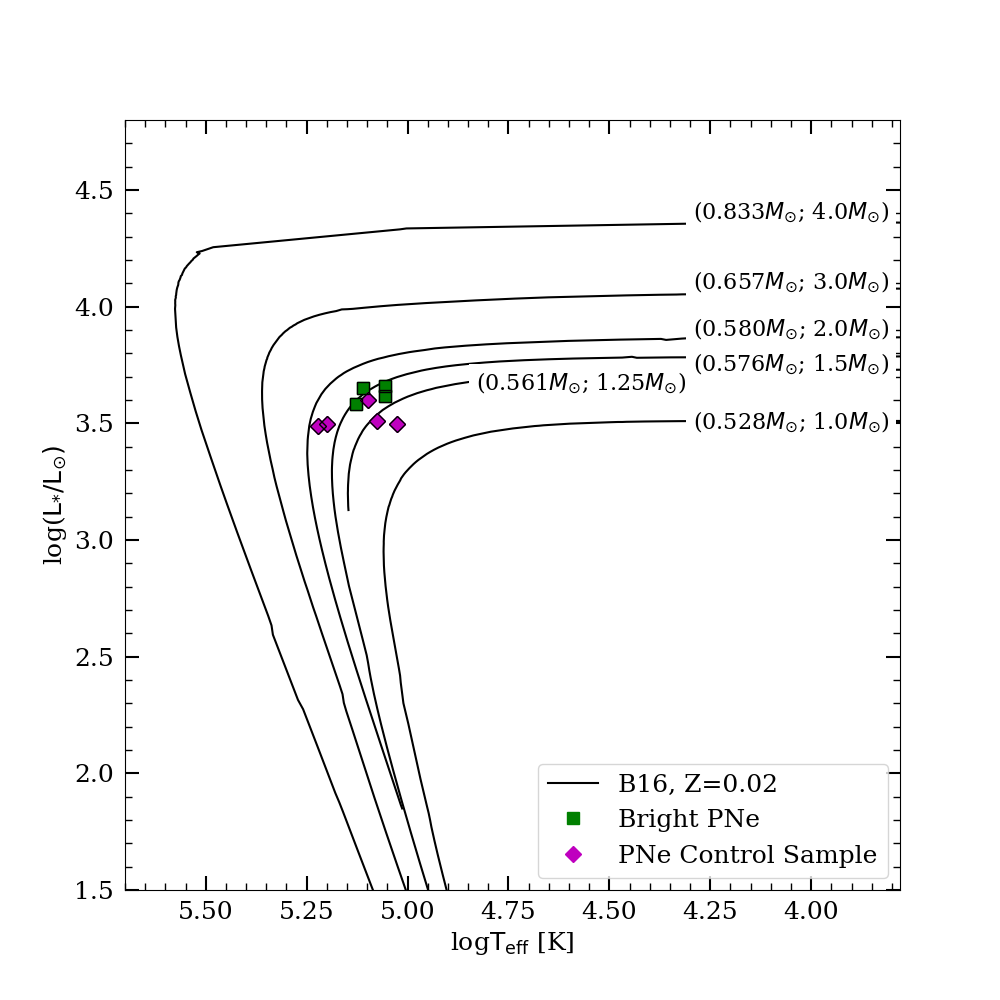}
    
        \caption{\label{Fig:HR_diagram_Dopita} {HR diagram  showing the location of the CSs of our PN samples as derived from Cloudy modeling in the \cite{Miller_Bertolami2016} model tracks. The final and corresponding initial masses are indicated on the tracks.     }}
\end{figure}

The positions in the HR diagram of our nine CSPNe are shown in Fig.~\ref{Fig:HR_diagram_Dopita}, together with post-AGB evolutionary tracks from 
\cite{Miller_Bertolami2016} for Z=0.02. Differences between these tracks and the classical tracks of \cite{Vassiliadis1994} -- basically, somewhat smaller final masses and substantially shorter, accelerated evolutionary times for the former -- have been discussed in recent papers \citep[cf.][]{Miller_Bertolami2016, Fang2018, Henry2018}.
Interpolation of the evolutionary tracks allows us to estimate $\mathrm{M_{f}}$ for each CSPN, and derive tentative masses $\mathrm{M_{i}}$ of their progenitors. The uncertainty associated with the interpolation is expected to be $\sim 10-15\%$ for accurately derived CSPN parameters  \citep{Serenelli2021}.

The resulting range of CSPN masses for the four brightest PNe in M31 (Table~\ref{table:m_E_T_L}) is remarkably narrow, $\mathrm{M_f}$=0.568 to 0.578 $\mathrm{M_{\odot}}$, with a mean value of $\mathrm{<M_f>}=$0.574$\pm$ 0.004 ~$\mathrm{M_{\odot}}$. That is in excellent agreement with the maximum final mass derived by \cite{valenzuela2019} from their PNLF simulations, $\mathrm{M_f}=$0.58~$\mathrm{M_{\odot}}$. The five PNe from the control sample show, in turn, a slightly broader range, $\mathrm{M_f}$=0.545 to 0.578 $\mathrm{M_{\odot}}$ with a mean value of $\mathrm{<M_f>}$=0.566$\pm$0.016$\mathrm{M_{\odot}}$.

This analysis shows that all nine PNe, including the brightest ones, originate from low-intermediate mass stars with $\mathrm{{M_{i}} < 2 M_{\odot}}$. 
The four brightest PNe (and also, to some extent, the less-bright control PNe) appear in Fig.~\ref{Fig:HR_diagram_Dopita} tightly clustered around the turning point of the tracks, close to the ``knee" where $\mathrm{T_{eff}}$ and $\mathrm{L_{*}}$ values simultaneously reach their maxima\footnote{It is remarkable that the 12 PNe from \cite{Kwitter2012} not in common with the current sample are also clustered near the same region of the HR diagram.}.

\begin{table}
        \caption{Luminosity and temperature of CSPNe from Cloudy modeling. The final and initial progenitor masses are derived from the \cite{Miller_Bertolami2016} evolutionary tracks.}      
        \label{table:m_E_T_L}
        \centering                          
        \begin{tabular}{l c c c c }      
                \hline\hline   
 PN &  $\mathrm{log(L_{*}/L_{\odot})}$ & $\mathrm{log(T_{eff})}$ &   $\mathrm{M_{f}(M_{\odot})}$ & $\mathrm{M_{i}(M_{\odot})}$ \\
\hline
\hline

M1687*   &   3.66  &   5.06  &  0.576 & 1.48 \\
M2068*   &   3.62  &   5.05  &  0.568 & 1.34 \\
M2538*   &   3.65  &   5.11  &  0.578 & 1.56 \\
M50*     &   3.59  &   5.13  &  0.573 & 1.42 \\
M1596    &   3.49  &   5.22  &  0.578 & 1.76 \\
M2471    &   3.50  &   5.20  &  0.578 & 1.70 \\
M2860    &   3.51  &   5.08  &  0.556 & 1.20 \\
M1074    &   3.50  &   5.03  &  0.545 & 1.12 \\
M1675    &   3.60  &   5.10  &  0.571 & 1.39 \\
\hline

        \end{tabular}
\end{table}

\section{Discussion}
\label{sec:discuss}

Understanding the [O~{\sc iii}] $\mathrm{\lambda5007}$ emission from the nebular shell as the CSPN quickly evolves across the HR diagram and, therefore, the origin of the observed PNLF is not trivial. The line emission strength strongly depends on the density, ionization, and chemical properties of the nebula, combined with the evolution of the post-AGB CS \citep{Schonberner2007}. In this section we analyze the properties of our targets trying to understand the nature of the brightest PNe as compared to the control sample, and to the predictions of PNLF theoretical models.

\subsection{Physical properties of the most luminous PNe and their central stars}
\label{sec:cspn}

Summarizing our observational results, we find that the nebular electron temperatures for the bright and control samples do not show significant differences, whereas electron densities appear larger for the former (Table~\ref{table:TemDen}). The $\mathrm{n_{e}}$ weighted average of the bright sample is $\sim$$40\%$ and $\sim$$18\%$ higher than in the control sample when measured from high- and low-ionization diagnostic lines, respectively. 

No significant differences are found for the masses and effective temperatures of the progenitor stars in the bright and control samples. They are in the range  $\mathrm{M_i=1.1}$ to $\mathrm{1.8~M_{\odot}}$, and $\mathrm{T_{eff}}=106\,000$ to $166\,000$~K.
All four bright PNe are located before the theoretical isochrone for 5000 a, while two PNe of the control sample appear slightly older, between 5000 and 10\,000 a. 

\cite{Marigo2004} studied the evolution of CSPNe by modeling the observed PNLF and its bright cutoff. In their models, the most luminous PNe are those originated from intermediate mass stars with final masses 
$\mathrm{M_{f}=0.7 \to 0.75~M_{\odot}}$, setting a lower limit for the inital mass $\mathrm{M_{i}>2.5~M_{\odot}}$. They also claimed that the maximum [O~{\sc iii}] $\lambda 5007$ emission would be attained by nebulae at the thick/thin transition during the CSPN post-AGB evolution.

On the other hand, \cite{Schonberner2007} presented a 1-D, spherically symmetric, radiative-hydrodynamics modeling of CSPNe with masses from $\mathrm{M_{f}}$=0.565 to 0.696
 $\mathrm{M_{\odot}}$. They concluded  that, at odds with \cite{Marigo2004}, 
the conversion of stellar UV radiation into line emission occurs most efficiently only for optically thick shells, attaining the observed PNLF [O~{\sc iii}] $\lambda 5007$ cutoff at M*$= -4.5$ mag for a stellar mass of $\mathrm{0.62~M_{\odot}}$, provided the PN remains optically thick. 

As discussed in \cite{Schonberner2007}, and more recently in \cite{valenzuela2019}, the question of the actual optical depth (thin or thick) to Lyman continuum photons for the brightest PNe is paramount to understand the PNLF, since the efficiency for transforming stellar UV photons into nebular ones is governed by the changes in optical depth during the coupled CS and PN evolution.
All nine of our PNe show relatively high electron densities and CS effective temperatures,
together with moderate values of the excitation parameter EC$\la 6$, pointing to optically thick PN shells. This is also supported by the presence of the [O~{\sc i}] $\mathrm{\lambda6300}$ line in all objects. Although the morphologies of our PNe are actually unknown, any departure from sphericity would also help to maintain the PNe opaque for longer during their evolution.
Three-dimensional modeling of PNe comprising dense equatorial tori and thin lobes is found to reproduce well the observed line emissivities including those of [O~{\sc iii}] \citep{Gesicki2016}.

Recently, \cite{Gesicki2018Nat} reevaluated the PNLF using the \cite{Miller_Bertolami2016} evolutionary tracks. Seven CSPNe tracks with $\mathrm{M_{i}=1.0-3.0~M_{\odot}}$ ($\mathrm{M_{f}=0.532-0.706~M_{\odot}}$), and the Torun photoionization code for the PN nebular emission were assumed. 
They showed that, for the five models with $\mathrm{M_{i}=1.1-2.5~M_{\odot}}$, the [O~{\sc iii}] emission reaches values of $\mathrm{m_{5007}}$ near the PNLF cutoff M* during hundreds to thousands of years, while the nebulae remain opaque. Assuming a Salpeter initial mass function and four star formation histories representative of different types of galaxies, synthetic PNLFs were built where the observed bright cutoff is shown to be populated by progenitors with moderate post-AGB stellar luminosities log$\mathrm{(L_{*}/L_{\odot})} = 3.75\pm 0.13$. This value is in agreement with, although slightly larger than, log$\mathrm{(L_{*}/L_{\odot})} = 3.63\pm 0.03$ found for the four brightest sources in our sample (Table~\ref{table:m_E_T_L}). Concerning the CSPN progenitor masses at the bright PNLF end, \cite{Gesicki2018Nat} estimate a range $\mathrm{M_{i}=1.1-2.0~M_{\odot}}$, which is similar to that of our full sample, $\mathrm{M_{i}=1.1-1.8~M_{\odot}}$. But we note that our four  brightest PNe display an even narrower range of masses, with a mean value $\mathrm{<M_i>=1.45\pm 0.09 ~M_{\odot}}$. According to \cite{Gesicki2018Nat}, such stars would remain above the bright-end cutoff M* for approximately one thousand years.

In summary, while the \cite{Marigo2004} and \cite{Schonberner2007} models predict CS masses substantially larger than what we estimated, the modeling by \cite{Gesicki2018Nat} using post-AGB tracks from \cite{Miller_Bertolami2016} agrees with our derived values for both luminosities and progenitor masses. 

For completion, it should be mentioned that \cite{Davis2018} studied 23 PNe in the bulge of M31 within 1 mag of the PNLF cutoff, finding substantial levels of extinction, c(H$\beta$)$\sim$0.6, most of which of circumstellar origin. This implies the presence of intrinsically overluminous objects, one-half to one magnitude brighter than the brightest PNe in our M31 disk sample. They would require, even using the evolutionary tracks from \cite{Miller_Bertolami2016}, CSPN masses $>$0.66 $\mathrm{M_\odot}$, that is, progenitors $>$2.5 $\mathrm{M_\odot}$, which are not expected to be found in a sufficient number in M31's bulge. \cite{Davis2018} suggest that objects populating the bright end of the PNLF in the bulge of M31 might not be PNe at all, but nebulae around symbiotic stars, as previously proposed by \cite{Soker2006}. Even if this limitation in the number of stars with $>$2.5 $\mathrm{M_\odot}$ would not apply in the disk of M31, the present study does not find any indication of such a high level of extinction for the target PNe. The c(H$\beta$) values range from 0 to 0.23 with, at most, $\sim$0.1 of circumstellar origin, resulting in the low CSPN masses calculated. We can also rule out in our sample indications of a symbiotic nature: all PNe are located well inside the PN area of the \cite{GM1995} [O~{\sc iii}] diagnostic diagram, and they show no signs of having bright and red CSs nor the high-excitation [O~{\sc vi}] line emission at 6825\AA\ typical of symbiotic stars \citep[cf.][]{mikolajewska2017}.

\subsection{ Large N/O ratios found in PNe with low-mass progenitors.}
\label{sec:largeN/O}

An intriguing question remains to be answered, namely, if progenitors of the brightest PNe are indeed $\mathrm{M_i\sim 1.5 ~M_{\odot}}$ stars, how can they produce the N overabundance found for at least five PNe in our sample (those with log(N/O)$\ga{-0.4}$ in Fig. \ref{Fig:NOvsHeO}), as well as for many PNe in previous studies \citep[cf.][]{Henry2018, Fang2018}. Our five sources show N/O values significantly larger than expected for their progenitor masses, predicted for $\mathrm{M_i\sim 3-3.5 ~M_{\odot}}$ stars according to models. 
Furthermore, data on our nine bright and control PNe in Fig. \ref{Fig:NOvsHeO} suggest a smooth progression toward larger N/O values as He/O decreases, in contrast with the abrupt increase in N/O at $\mathrm{M_i\sim 3.5-4 ~M_{\odot}}$ predicted by the models.

The issue is related to the low-end mass limit for stars experiencing hot bottom burning (HBB) during the AGB. Some models set the enrichment threshold at $\mathrm{M_{i}\ga{4}~M_{\odot}}$ \citep{Karakas2016, Cristallo2015, Marigo2017} while other models that include overshooting at the boundary core \citep {Miller_Bertolami2016, Ventura2018}, decrease it to $\mathrm{M_{i}\sim{3}~M_{\odot}}$. The limit has been continuously pushed down by the observations: \cite{Davis2019} found a 3.4$\mathrm{~M_{\odot}}$ CS of a Type I PN in an open cluster in M31 where indeed HBB has taken place, whereas samples of PNe from the Milky Way \citep{Henry2018} and M31 \citep{Fang2018} suggest the limit could be as low as $\mathrm{M_{i}\sim{2}~M_{\odot}}$. Our data indicate even lower masses, $\mathrm{M_{i}\la{1.8}~M_{\odot}}$, for PNe with the larger values of N/O in our sample, and no extant model is able, to our knowledge, to reach near that value.

A possible way to circumvent this conundrum is to reevaluate the IFMR of low-intermediate mass stars, and a recent paper from \cite{Marigo2020} might represent an important step forward in that direction. They carefully determine initial and final masses for a sample of white dwarfs (WDs) in old Galactic open clusters, and find a kink at $\mathrm{M_{i}= 1.65-2.10~M_{\odot}}$ in the IFMR, whose peak at $\mathrm{M_{i}\sim 1.8-1.9~M_{\odot}}$ corresponds to measured WD masses of $\mathrm{M_{f}= 0.70-0.75~M_{\odot}}$. It is important to emphasize that those semiempirical data strongly disagree with current models of AGB evolution: CSPN with masses of $\mathrm{M_{f}= 0.70-0.75~M_{\odot}}$ would originate from main sequence stars with much larger masses $\mathrm{M_{i}\ga 3-3.5~M_{\odot}}$ according to the \cite{Vassiliadis1994} or \cite{Miller_Bertolami2016} models. Moreover, stars with masses around $\mathrm{M_{i}\sim 1.5~M_{\odot}}$, located at the lower-mass side of the IFMR kink, would evolve into relatively high-mass, $\mathrm{M_{f}\sim 0.6~M_{\odot}}$, WDs according to \cite{Marigo2020}. That is larger than \cite{Miller_Bertolami2016} model tracks predict (a $\mathrm{M_{f}\sim 0.6~M_{\odot}}$ star would originate from a $\mathrm{M_{i}\sim 2.3~M_{\odot}}$ star according to his modeling), although we note that $\mathrm{M_{i}= 1.5~M_{\odot}}$ progenitors do end as $\mathrm{M_{f}\sim 0.6~M_{\odot}}$ dwarfs in \cite{Vassiliadis1994} tracks. 

Determining accurate masses from WD stars and their progenitors is a difficult task, and the $\mathrm{M_{i}\sim 1.5~M_{\odot}}$ region of the IFMR from \cite{Marigo2020} is really based on data from a single open cluster, M67 (their Fig. 1). A recent reevaluation of seven WD masses belonging to that cluster by \cite{Canton2021} confirm the mean value of $\mathrm{M_{f}= 0.60\pm 0.01~M_{\odot}}$ for the WDs, and $\mathrm{M_{i}= 1.52\pm 0.04~M_{\odot}}$ for their progenitors (quoted uncertainties are formal errors of the mean, not dispersion of the sample). 

Assuming that $\mathrm{M_{i}\sim 1.5~M_{\odot}}$ stars, like the proposed progenitors of our four brightest PNe, would end as $\mathrm{M_{f}\sim 0.6~M_{\odot}}$ CSPN, we wish to determine what effects this would have on the nucleosynthesis and the yields of elements such as He, C, N, and possibly O. We also want to know if the discrepancy between predicted and observed N/O ratios will be solved or alleviated, given the now more massive stellar cores of the AGB stars.

In the mass regime $\mathrm{M_{i}\sim 1.5-2.0~M_{\odot}}$ a process takes place that is key to understanding the AGB and post-AGB evolution: stars with solar metallicity and masses up to $\mathrm{M_{i}\sim1.8-1.9~M_{\odot}}$ experience a core He flash at the beginning of their core He-burning phase, while those with $\mathrm{M_{i}=2.0~M_{\odot}}$ and larger do not. That defines the fundamental frontier between low- and intermediate-mass stars, and changes not only the evolutionary timescales, but also the strength of the 3rd dredge-up and the chemistry. As mentioned earlier, that mass limit at $\mathrm{M_{i}\sim1.8-1.9~M_{\odot}}$ also coincides with the observed kink in the IFMR, making CSPN and WD stars that are more massive than previously thought \citep{Marigo2020}. On the other hand, according to \cite{Miller_Bertolami2016} modeling, it is just at $\mathrm{M_{i}=1.5~M_{\odot}}$ where the largest carbon enrichment is expected, almost double that of a $\mathrm{M_{i}=2.0~M_{\odot}}$ star (his Fig. 6). If such an $\mathrm{M_{i}=1.5~M_{\odot}}$ star were to end as an $\mathrm{M_{f}=0.60~M_{\odot}}$ CSPN its properties would be quite different from one with $\mathrm{M_{f}=0.576-0.580~M_{\odot}}$ in terms of 3rd dredge-up efficiency, initial H envelope, and H-burning rate, and it would have higher luminosity \citep{Miller_Bertolami2016}. However, in terms of the N surface enrichment, only at masses $\mathrm{M_{i}>3.0~M_{\odot}}$ does there appear a significant jump upward according to all extant models, and therefore the origin of the disagreement between modeled and observed N/O ratios remains unexplained.

Further efforts to improve low-mass star models, including extra-mixing processes beyond overshooting, such as rotation (meridional circulation and diffusion by shear turbulence), magnetic field, or thermohaline mixing, have been encouraged by better and better observational data in recent years \citep{Delgado_Inglada2015, Henry2018, Fang2018}. The substantial N overabundance measured in several of the most luminous PNe in M31 provides motivation for these continuing efforts.

\section{Conclusions}

The PNLF is an astrophysical observable that appears to have a universal bright cutoff. This is thought to be caused by an unexpected constancy in the combined results from the complex physics of stellar interiors, stellar evolution all along the HR diagram, nucleosynthesis and dredge-up processes, the physics of envelope ejection, ionization, and the dynamical evolution of the PNe.
In addition to having a mild dependence on metallicity, all these processes seem to ``conspire" to yield an extremely simple result: All the brightest PNe in old and younger stellar systems have a unique cutoff magnitude at $\mathrm{M^{*} = -4.54\pm0.05}$.

While the building of the PNLF itself has been recently explained using updated stellar evolutionary tracks (\citealt{Gesicki2018Nat}; but see \citealt{Davis2018}), the detailed nature of the PNe populating the bright end of the PNLF is still largely unknown due to distance uncertainties for Galactic PNe and the inherent faintness of PNe in external galaxies.

The goal of this work has been to analyze in detail a representative sample of bright PNe in the galaxy M31, our closest Milky Way-like neighbor, which hosts the largest collection of PNe known in any galaxy. We carefully selected and observed two samples of ``bright PNe" (four objects inside the bright cutoff bin) and ``control PNe" (five objects typically 0.5 mag fainter). We supplemented them with an ``extended sample" of another 21 PNe from previous works observed with the same telescope and instrument and reanalyzed here to ensure homogeneity.
Relevant conclusions from this work are:
\begin{itemize}
\item{10.4m~GTC spectra obtained under excellent observing conditions allow us to reach the unprecedented depth of a 3$\sigma$ rms signal at $\sim$0.2$\%$ of I($H\beta$) in the blue, even in the faintest nebulae.}
\item{These spectra are carefully and homogeneously analyzed  using updated physics and the most recent ICFs for abundance calculations, minimizing the uncertainties in the derived physico-chemical properties.}
\item{The four brightest and the five control PNe show a remarkable uniformity in nebular characteristics: Electron temperatures range between T$_{e}$([O~{\sc iii}])=10\,100 and 12\,700~K. All nebulae show large electron densities, $\mathrm{n_{e}([Ar~{\sc IV}])}=5000-32000$ $\mathrm{cm^{-3}}$, those of the brightest nebulae being, on average, $\sim40\%$ larger than those of the control sample. The nebular ECs also span a narrow range of EC=4.5-5.6 for the bright and control nebulae.}
\item{Derived abundances for He, Ne, N, O, Ar, and S agree neatly with their predicted qualitative interrelations, and this is the first time such global agreement is observed in any PN sample, including those in the Milky Way. Namely, {\it i)} three observed alpha-elements, Ne, Ar, and S, behave, as theoretically expected, in lockstep with respect to O, and the expected tight S/H versus O/H correlation is measured for the first time (while the sulfur anomaly persists)}. {\it ii)} He and N appear strongly correlated with each other and with O, following their expected dependence on metallicity but, significantly, only for the bright and control, not the extended, samples. {\it iii)} The expected He versus O, N/O versus He, and N/O versus O correlations are observed clearly for the first time, again only for the bright and control samples. {\it iv)} The diagram N/O versus He/O is introduced and shows a tight anticorrelation (once more, only for the bright and control samples), allowing comparison with two families of models, MONASH  \citep{Karakas2016} and ATON \citep{Ventura2018}. Whereas He/O ratios are consistent with model results, measured N/O values are 1.5-3 times larger than model predictions for the range of progenitor masses derived for five of our PNe. This, adding to previous cases from the literature, indicates an important limitation in the models. {\it v)} The contrasting behavior of the bright and control samples, on the one hand, and the extended sample, on the other, is thought to be caused by the fact that  the extended sample is an inhomogeneous collection of PNe that belong to different M31 components (the extended halo or its outer substructures, the Warp, G1 Clump, Northern Spur, and Giant Stream), with different chemical compositions and star formation histories.  
\item{A primary goal of this work was to determine the CS masses (and progenitor initial masses via an IFMR) of these bright PNe. Cloudy modeling indicates a remarkably homogeneous set of luminosities and effective temperatures, with average values and a 1$\sigma$ dispersion of
$\mathrm{<L_{*}/L_{\odot}>} = 4\,300\pm 310$, $\mathrm{T_{eff}} = 122\,000\pm 10\,600$ K for the CS of the bright sample, and 
$\mathrm{<L_{*}/L_{\odot}>} = 3\,300\pm 370$, $\mathrm{T_{eff}} = 135\,000\pm 26\,000$ K for those in the control sample. 
}
\item{The CS of the four brightest PNe appear closely grouped at the ``knee" of the $\mathrm{M_i=1.5~M_{\odot}}$ evolutionary track from \cite{Miller_Bertolami2016}, whereas those of the five control PNe span a slightly broader range of initial masses, $\mathrm{M_i=1.1 \to 1.8~M_{\odot}}$
}
\end{itemize}

Our overall conclusion is that the brightest PNe in the (extended) disk of M31 are normal, moderately dense nebulae, with low-mass, $\mathrm{M_i\sim 1.5~M_{\odot}}$, progenitors, that, according to the modeling by \cite{Gesicki2018Nat}, are able to populate the bright end of the PNLF for $\sim$1\,000 years. We note that the stellar mass distribution in a star-forming environment such as the disk of M31 spans a large range, including stars with higher masses that reach larger luminosities in their post-AGB evolution. We plan to extend this work to stellar systems with different star formation histories or metallicities, starting from the Large Magellanic Cloud, to gain more comprehensive observational constraints about PNe at the tip of the PNLF.

\begin{acknowledgements}
During the preparation of this article, while finishing her PhD at the Instituto de Astrofísica de Canarias, Rebeca Galera-Rosillo untimely passed away at the age of 31. That had a devastating impact on her family, colleagues and friends. Her passion for astrophysics, her solidarity with the underprivileged and her never-ending smile will not be forgotten. \\
AM, and RC heartily acknowledge the material, and above all, personal support provided by BB, KK, LM, and EV to Rebeca during her stays at their home institutions.
Particular thanks are due to Christophe Morisset for his help with PyCloudy and Machine Learning programs to constrain the ICFs, and to Simone Madonna for many discussions and continued support to Rebeca.
This work is based on observations made with the Gran Telescopio Canarias (GTC), installed in the Spanish Observatorio del Roque de los Muchachos of the Instituto de Astrofísica de Canarias, in the island of La Palma.
AM, RC, JG-R, and DJ acknowledge support from the State Research Agency (AEI) of the Spanish Ministry of Science, Innovation and Universities (MCIU) and the European Regional Development Fund (FEDER) under grant AYA2017-83383-P, and support under grant P/308614 financed by funds transferred from the Spanish Ministry of Science, Innovation and Universities, charged to the General State Budgets and with funds transferred from the General Budgets of the Autonomous Community of the Canary Islands by the MCIU. JG-R acknowledges support from the Severo-Ochoa excellence programs, SEV-2015-0548 and CEX2019-000920-S, DJ also acknowledges support from the Erasmus+ program of the European Union under grant number 2020-1-CZ01-KA203-078200, and EV support from Spanish grant PGC2018-101950-B-100.  
\end{acknowledgements}

\bibliographystyle{aa} 
\bibliography{41890.bib} 

\begin{appendix}

\section{Supporting material}
\label{sec:sup_mat}

\begin{figure*}
	\centering
	\subfigure {\includegraphics[width=8cm]{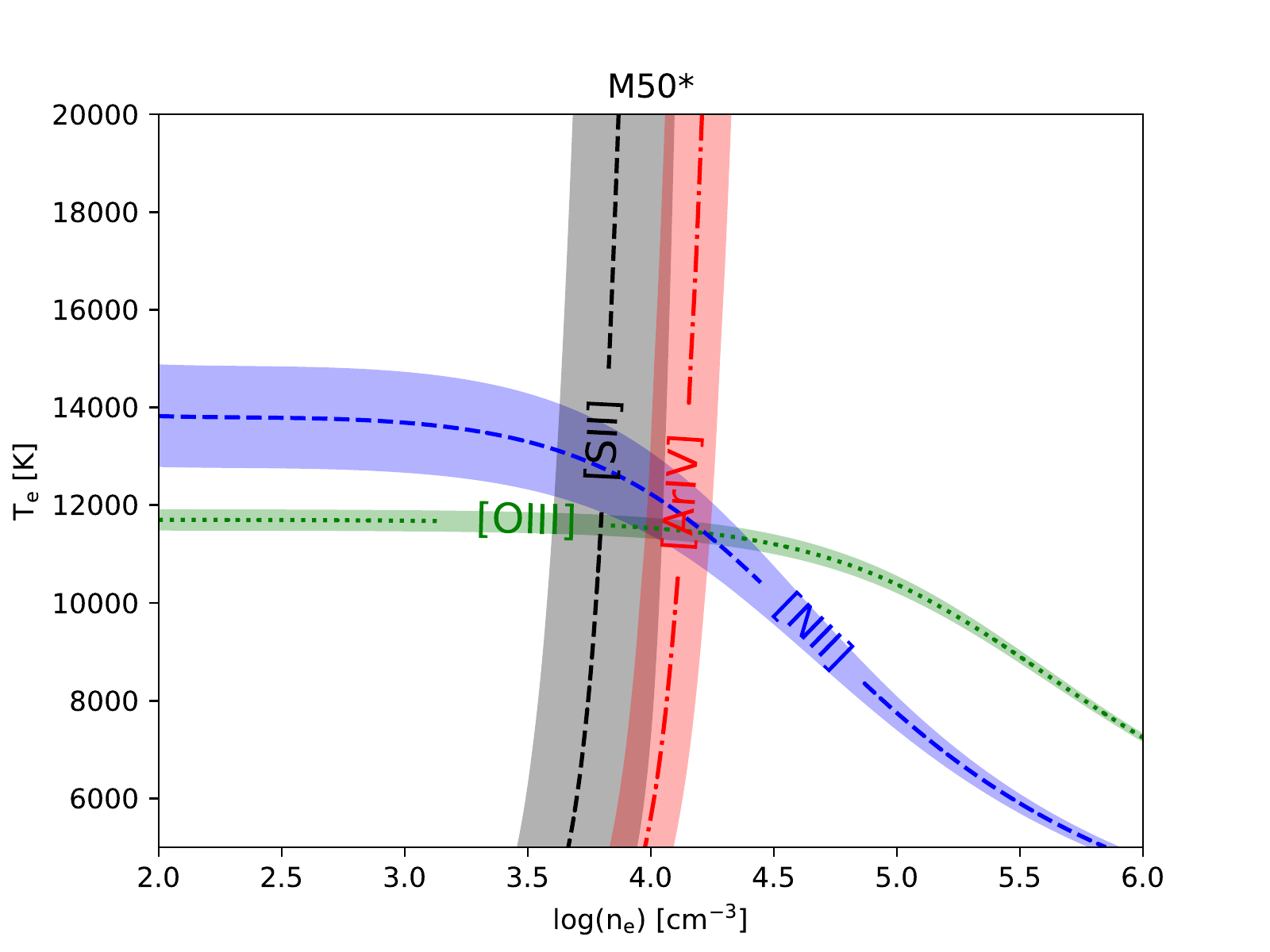}}
	\subfigure {\includegraphics[width=8cm]{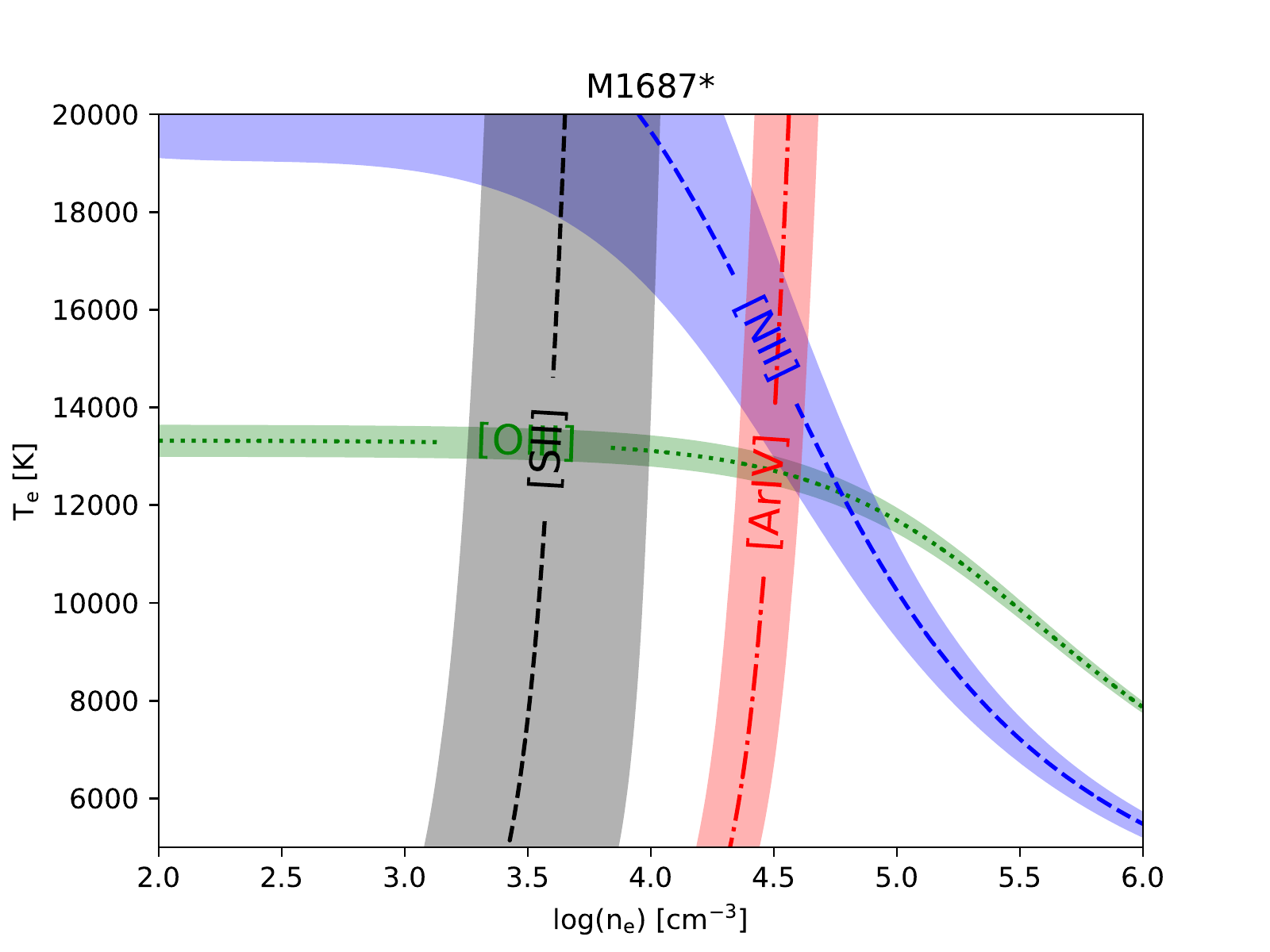}}
	\subfigure {\includegraphics[width=8cm]{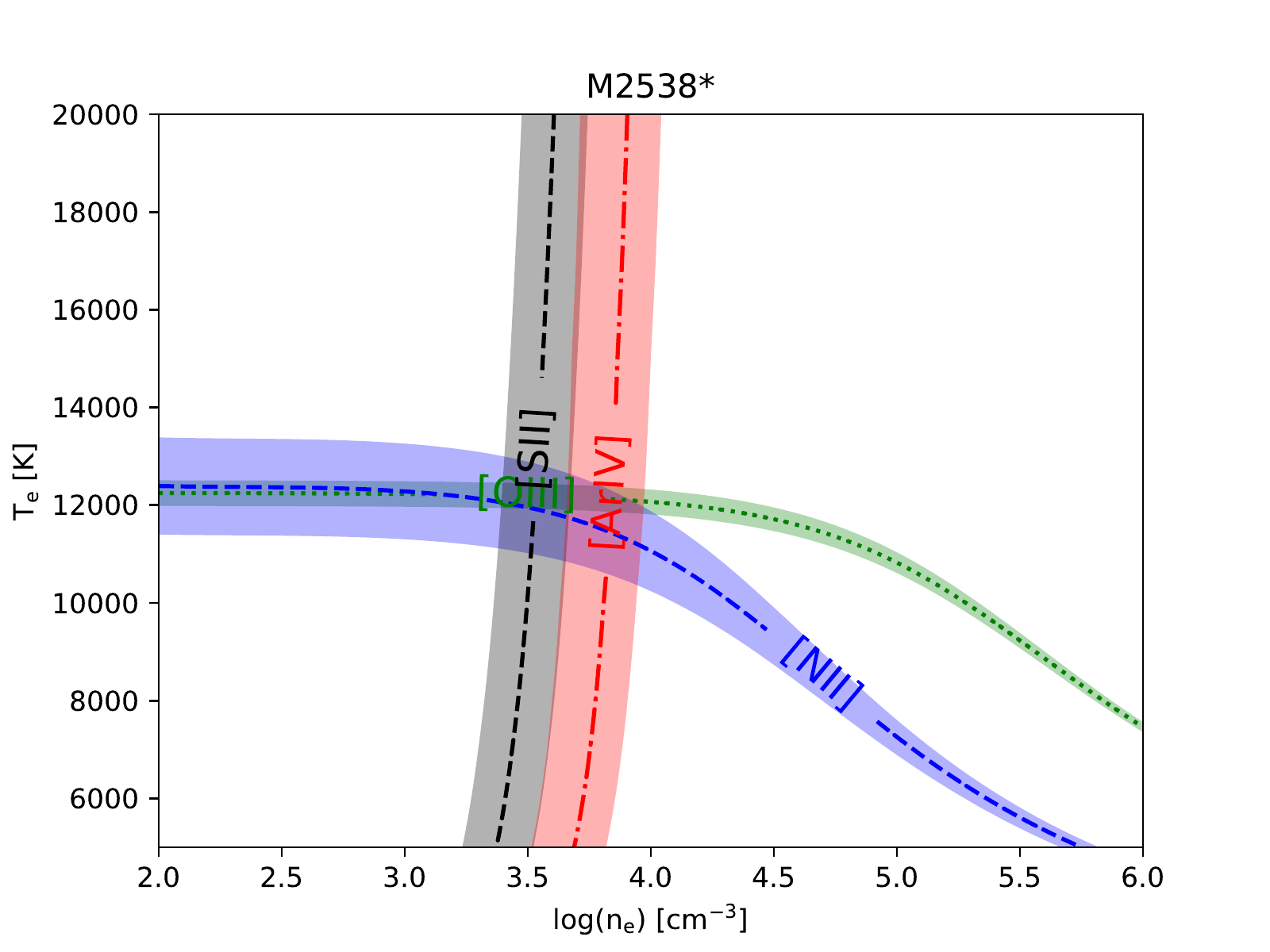}}
	\subfigure {\includegraphics[width=8cm]{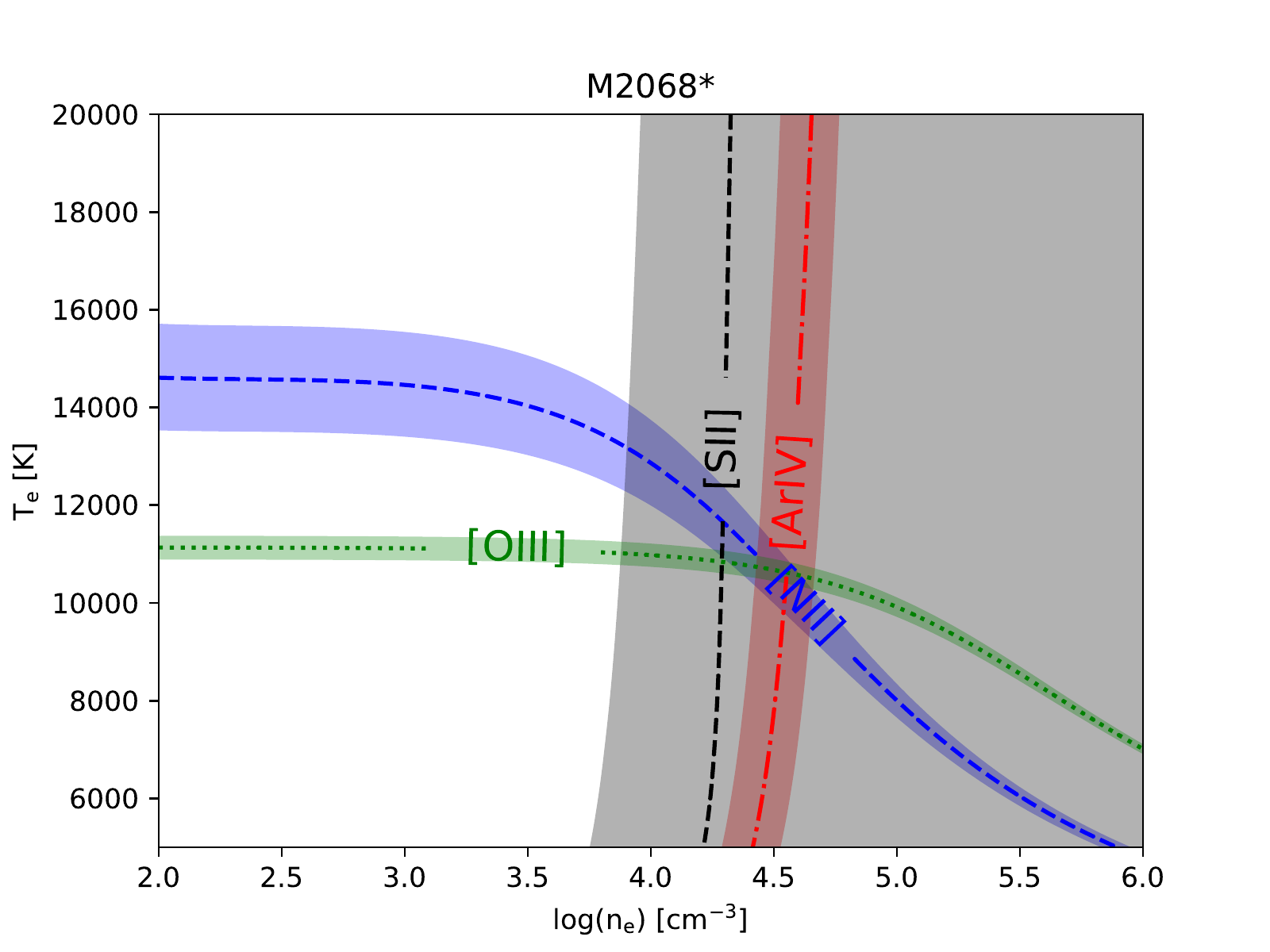}}
	\caption{\label{Fig:plasma_diag_b}  Plasma diagnostics diagrams of the observed PNe bright tip sample. } 
\end{figure*}

\begin{figure*}
	\centering
	\subfigure {\includegraphics[width=8cm]{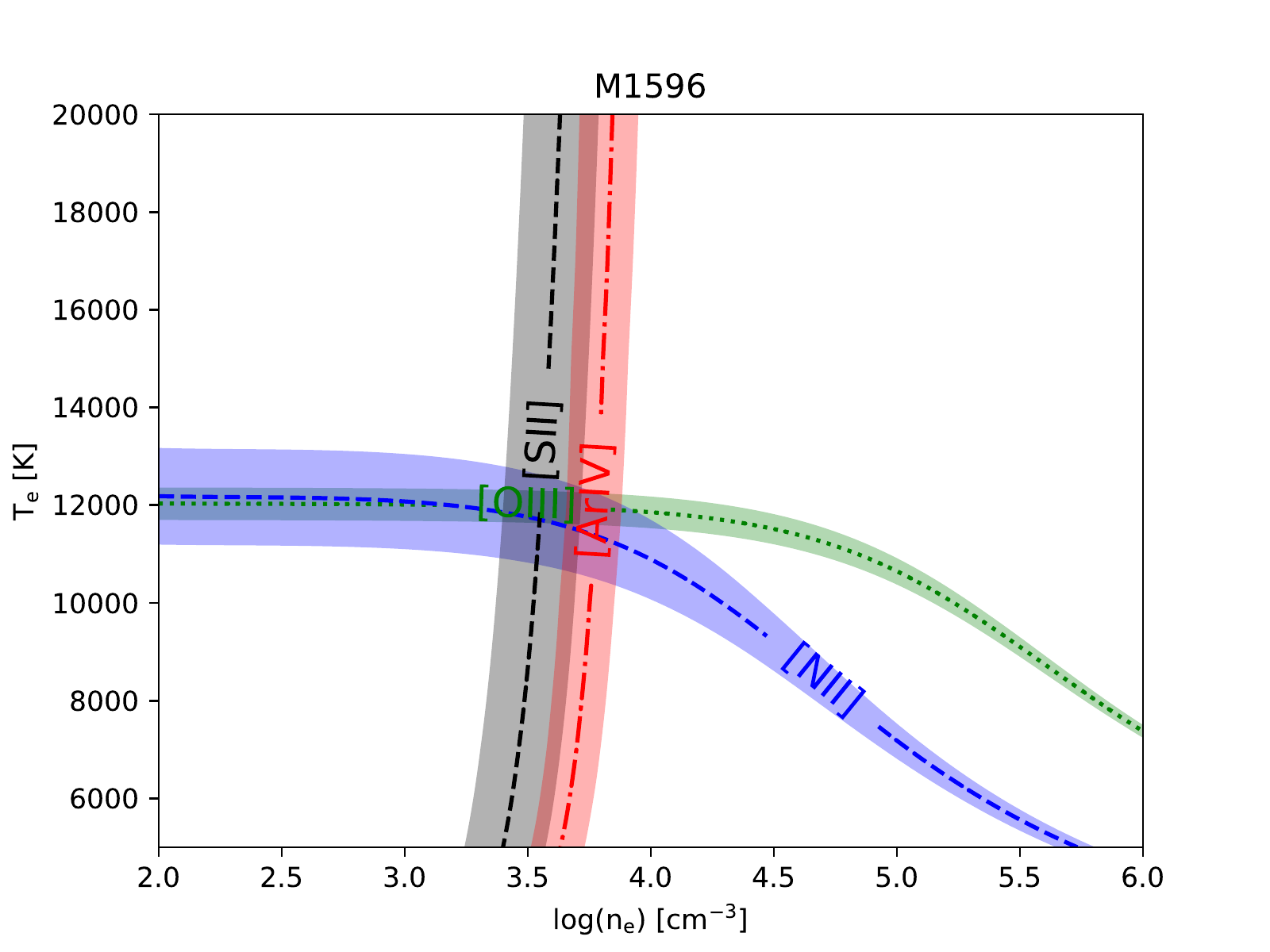}}
	\subfigure {\includegraphics[width=8cm]{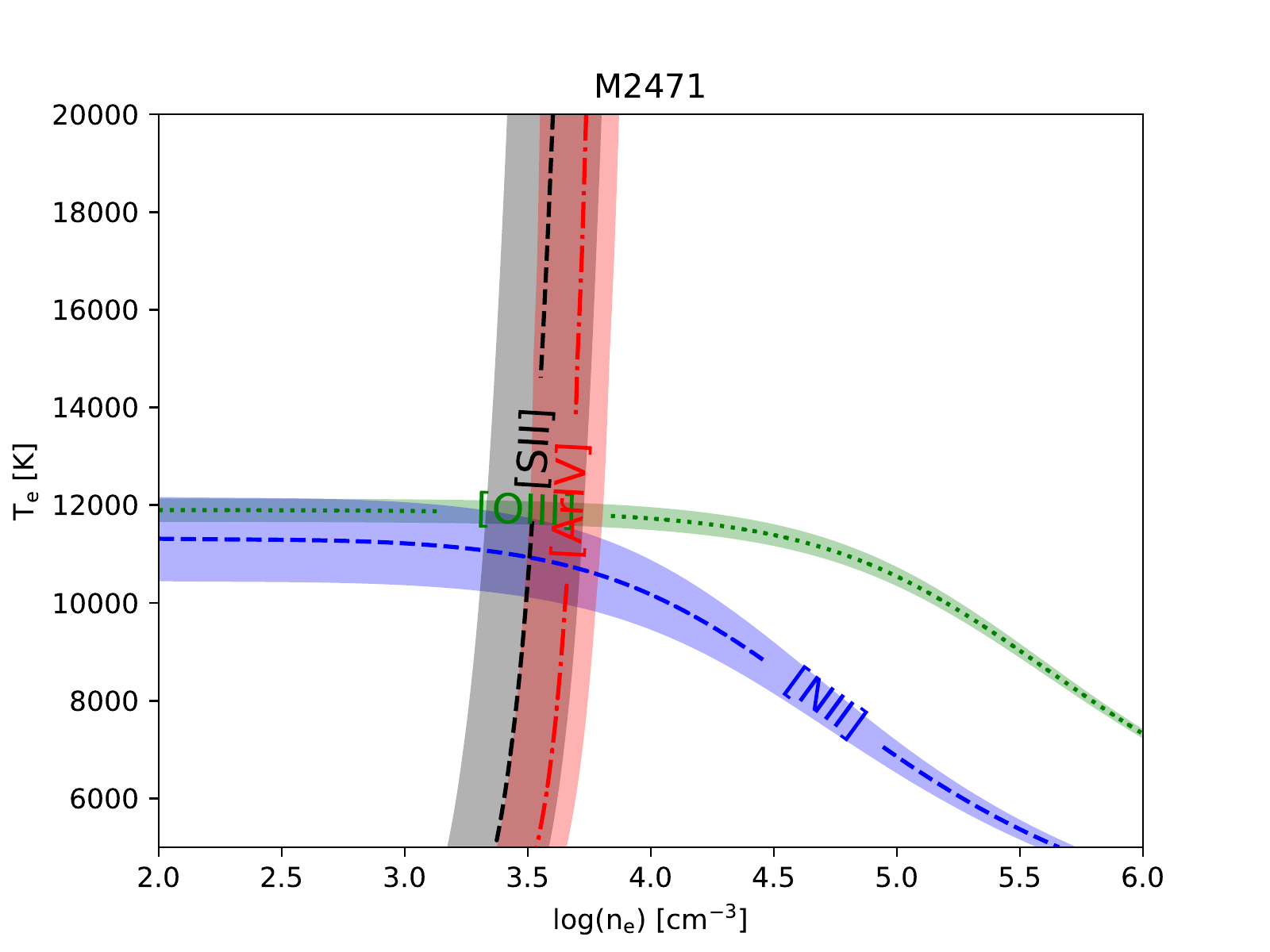}}
	\subfigure {\includegraphics[width=8cm]{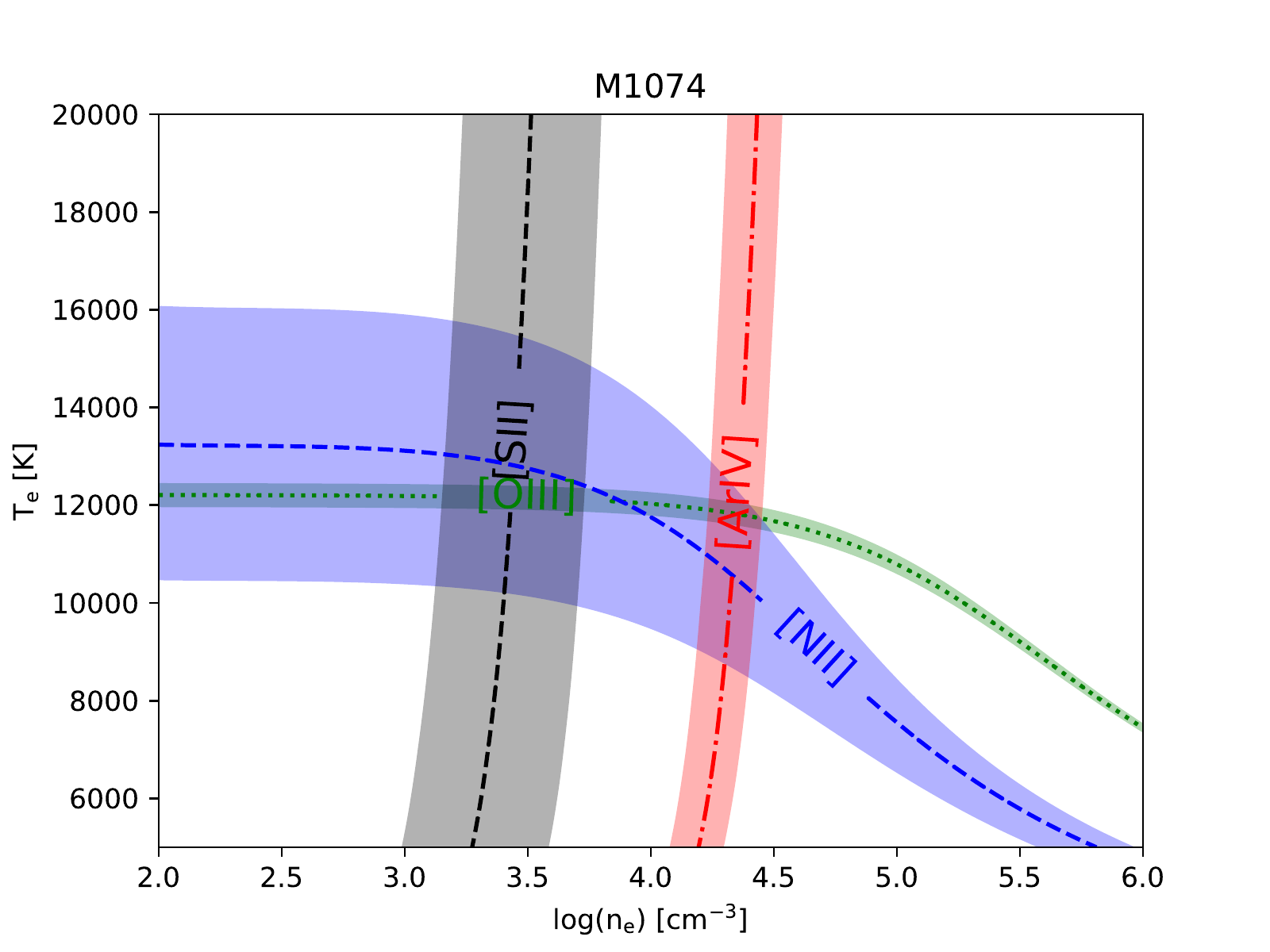}}
	\subfigure {\includegraphics[width=8cm]{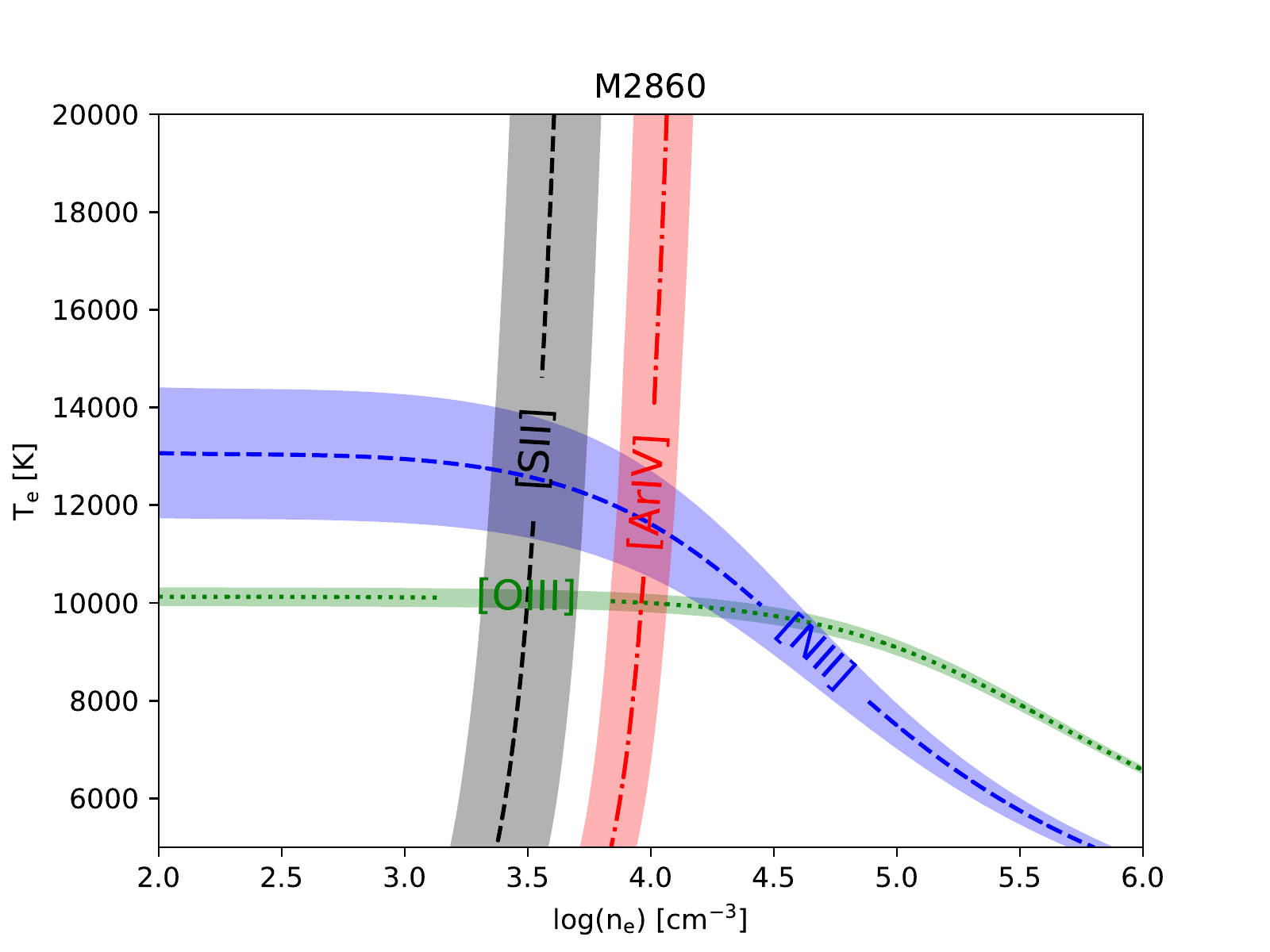}}
	\subfigure {\includegraphics[width=8cm]{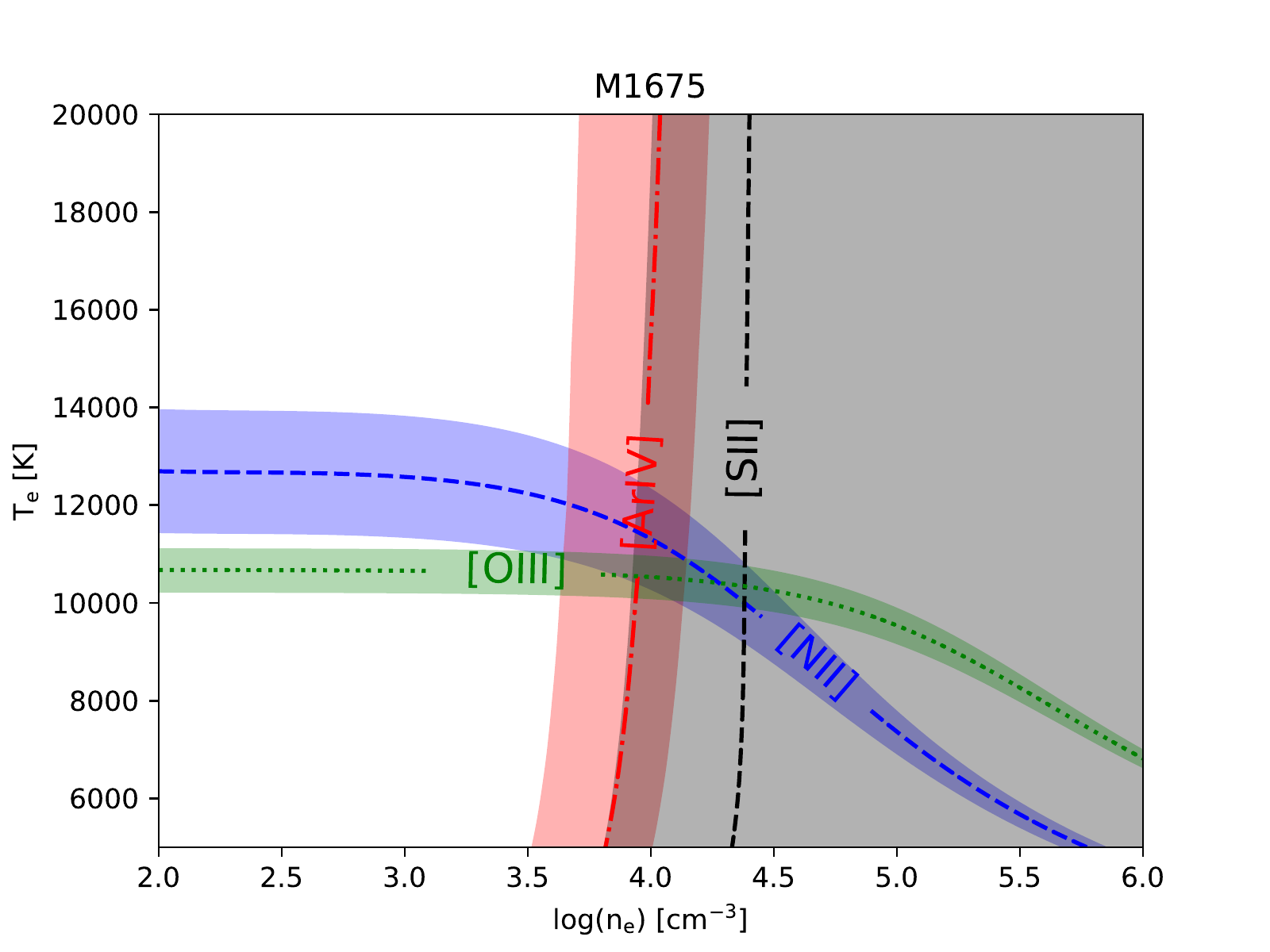}}
	\caption{\label{Fig:plasma_diag_cs} Plasma diagnostics diagrams of the observed PNe control sample.} 
\end{figure*}

In this appendix we include the following material:
plasma diagnostics for the four PNe in the bright sample (Fig.~\ref{Fig:plasma_diag_b}); plasma diagnostics for the five PNe in the control sample (Fig.~\ref{Fig:plasma_diag_cs}); line identification and flux table for all nine PNe in our bright and control samples (Table~\ref{table:fluxes_one_table}); and ionic abundances (X$^{i+}$/H$^+$) for the PNe in the bright and control samples (Table~\ref{table:ionic_abundances}).
 
\clearpage
\newpage

\onecolumn
\begin{longtable}{lc@{\hspace{6pt}} r@{\hspace{0pt}} c@{\hspace{0pt}} l c@{\hspace{6pt}} r@{\hspace{0pt}} c@{\hspace{0pt}} l c@{\hspace{6pt}} r@{\hspace{0pt}} c@{\hspace{0pt}} l c@{\hspace{6pt}} r@{\hspace{0pt}} c@{\hspace{0pt}} l c@{\hspace{6pt}} r@{\hspace{0pt}} c@{\hspace{0pt}} l }
        \caption{Measured (F) and de-reddened (I) lines fluxes.}         \label{table:fluxes_one_table}  \\
\hline
\multicolumn{1}{c}{}        
  &   \multicolumn{4}{c}{M1687*}    & \multicolumn{4}{c}{M2068*}  &  \multicolumn{4}{c}{M2538*}  &  \multicolumn{4}{c}{M50*} \\
\hline

 \multicolumn{1}{c}{Ident.} &   \multicolumn{1}{c}{F}  & \multicolumn{3}{c}{I $\pm$ $\Delta$I}
 & \multicolumn{1}{c}{F} &  \multicolumn{3}{c}{ I $\pm$ $\Delta$I} &  \multicolumn{1}{c}{F} & \multicolumn{3}{c}{I $\pm$ $\Delta$I} &  \multicolumn{1}{c}{F} & \multicolumn{3}{c}{I $\pm$ $\Delta$I}  \\
\hline
\hline

$\textrm{[O~{\sc ii}]}$ 3727  &    17.6   &  20.3   &   $\pm$ & 1.4   &     38.0   &  41.9   &   $\pm$ & 2.4   &   50.4 &  50.7   &   $\pm$ & 1.6    &   43.1  &  50.2    &   $\pm$ & 3.3     \\                  
H10 3798    &    --     &         &   --    &       &      --    &         &   --    &       &   3.39 &  3.41   &   $\pm$ & 0.63   &   4.72  &  5.45    &   $\pm$ & 1.59    \\             
H9 3835    &    --     &         &   --    &       &      4.70  &  5.15   &   $\pm$ & 0.66  &   6.92 &  6.96   &   $\pm$ & 0.55   &   4.74  &  5.46    &   $\pm$ & 1.48     \\             
$\textrm{[Ne~{\sc iii}]}$ 3868    &    147    &  167    &   $\pm$ & 9     &     126    &  138    &   $\pm$ & 7.08  &  93.1  &  93.7   &   $\pm$ & 2.8    &   104   &  119     &   $\pm$ & 6        \\             
He~{\sc i}+H8 3889    &    12.8   &  14.6   &   $\pm$ & 1.1  &     12.8    &  14.0   &   $\pm$ & 1.01  &  13.6  &  13.7   &   $\pm$ & 0.7    &   13.4  &  15.4    &   $\pm$ & 1.9      \\             
$\textrm{[Ne~{\sc iii}]}$ 3967\tablefootmark{a}    &    58.9   &  49.1   &   $\pm$ & 3.41  &     53.3   &  40.5   &   $\pm$ & 2.90  &   33.8 &  16.8   &   $\pm$ & 1.3    &   42.8  &  31.2    &   $\pm$ & 3.0      \\   
He~{\sc i} 4026    &    5.31   &  5.95   &   $\pm$ & 1.20  &      2.28  &  2.46   &   $\pm$ & 0.91  &  3.07  &  3.09   &   $\pm$ & 0.68   &   2.14  &  2.41    &   $\pm$ & 1.64      \\             
$\textrm{[S~{\sc ii}]}$ 4068    &    --     &         &    --   &       &      3.49  &  3.76   &   $\pm$ & 0.77  &  --    &         &   --    &        &   3.36  &  3.77    &   $\pm$ & 2.07      \\             
$\textrm{[S~{\sc ii}]}$ 4071    &    --     &         &    --   &       &      2.36  &  2.54   &   $\pm$ & 0.90  &  --    &         &   --    &        &   0.91  &  1.02    &   $\pm$ & 2.28      \\             
H6 4101\tablefootmark{b}    &    25.8   &  28.6   &   $\pm$ & 1.62  &     24.8   &  26.6   &   $\pm$ & 1.39  &  24.9  &  24.9   &   $\pm$ & 1.0    &   23.4  &  25.9    &   $\pm$ & 2.4      \\             
H5 4340\tablefootmark{c}    &    44.5   &  47.8   &   $\pm$ & 2.01  &     43.7   &  45.8   &   $\pm$ & 1.81  &  44.5  &  44.5   &   $\pm$ & 1.4    &   41.8  &  44.8    &   $\pm$ & 1.7      \\             
$\textrm{[O~{\sc iii}]}$ 4363    &    24.8   &  26.6   &   $\pm$ & 1.39  &     13.4   &  14.0   &   $\pm$ & 0.80  &  16.8  &  16.9   &   $\pm$ & 0.8    &   16.0  &  17.3    &   $\pm$ & 0.7      \\             
$\textrm{Ne~{\sc ii}}$  4428    &     --    &         &   --    &       &      --    &         &    --   &       &  0.63  &  0.63   &   $\pm$ & 0.15   &   0.91  &  0.97    &   $\pm$ & 0.37      \\             
He~{\sc i} 4472    &    5.63   &  5.94   &   $\pm$ & 0.49  &      5.47  &  5.67   &   $\pm$ & 0.33  &  4.22  &  4.24   &   $\pm$ & 0.27   &   4.71  &  4.99    &   $\pm$ & 0.36      \\             
He~{\sc ii} 4542    &     --    &         &    --   &       &      --    &         &    --   &       &  --    &         &   --    &        &     --  &        &    --   &           \\ 
N~{\sc iii}+O~{\sc ii} 4640    &    --     &         &    --   &       &      0.93  &  0.95   &   $\pm$ & 0.85  &  3.15  &  3.17   &   $\pm$ & 0.59   &   1.25  &  1.30    &   $\pm$ & 0.38      \\              
O~{\sc ii} 4650    &    --     &         &    --   &       &      --    &         &    --   &       &  --    &         &   --    &        &   1.71  &  1.77    &   $\pm$ & 0.37      \\             
He~{\sc ii} 4686    &    2.82   &  2.89   &   $\pm$ & 0.48  &      3.74  &  3.80   &   $\pm$ & 0.25  &  11.9  &  12.0   &   $\pm$ & 0.4    &   14.0  &  14.4    &   $\pm$ & 0.5      \\             
$\textrm{[Ar~{\sc iv}]}$ 4711\tablefootmark{d}    &    2.68   &  1.87   &   $\pm$ & 0.40  &      2.11  &  1.34   &   $\pm$ & 0.27  &  3.23  &  2.53   &   $\pm$ & 0.28   &   3.08  &  2.40    &   $\pm$ & 0.41      \\      
$\textrm{[Ar~{\sc iv}]}$ 4740    &    5.17   &  5.26   &   $\pm$ & 0.35  &      3.93  &  3.98   &   $\pm$ & 0.29  &  3.32  &  3.34   &   $\pm$ & 0.38   &   4.25  &  4.35    &   $\pm$ & 0.35      \\             
H$\beta$ 4861\tablefootmark{e}    &    100    &  100    &   $\pm$ & 3     &    100     &  100    &   $\pm$ &  3    & 100    &  100    &   $\pm$ & 3      &   100   &  100     &   $\pm$ & 3        \\             
He~{\sc i} 4922    &    0.86   &  0.86   &   $\pm$ & 0.43  &      1.42  &  1.42   &   $\pm$ & 0.39  &  1.05  &  1.06   &   $\pm$ & 0.30   &   0.75  &  0.75    &   $\pm$ & 0.35      \\             
$\textrm{[O~{\sc iii}]}$ 4959    &    608    &  602    &   $\pm$ & 18    &    517.2   &  514.2  &   $\pm$ & 15.5  & 471    &  474    &   $\pm$ & 14     &   552   &  550     &   $\pm$ & 17      \\              
$\textrm{[O~{\sc iii}]}$ 5007    &  1812$^{f}$ & 1794    & $\pm$   & 54    &   1541$^{f}$ & 1532    & $\pm$  & 42     & 1404$^{f}$ & 1412    &   $\pm$ & 38   & 1646$^{f}$ & 1639     &   $\pm$ & 41       \\             
$\textrm{[NI]}$ 5198    &    --     &         &    --   &       &     --     &         &   --   &        &  --    &         &  --     &        &   0.51  &  0.49    &   $\pm$ & 0.33       \\             
He~{\sc ii} 5411    &    --     &         &    --   &       &     --     &         &   --   &        &  1.54  &  1.55   &   $\pm$ & 0.30   &   1.47  &  1.40    &   $\pm$ & 0.22       \\             
$\textrm{[Cl~{\sc iii}]}$ 5518    &    --     &         &    --   &       &     --     &         &   --   &        &   --   &         &   --    &        &   0.41  &  0.39    &   $\pm$ & 0.18       \\             
$\textrm{[Cl~{\sc iii}]}$ 5538    &    --     &         &    --   &       &     --     &         &   --   &        &   --   &         &   --    &        &   0.84  &  0.79    &   $\pm$ & 0.18       \\            
$\textrm{[N~{\sc ii}]}$ 5755    &    2.20   &  2.03   &   $\pm$ & 0.41  &      2.64  &  2.50   &   $\pm$ & 0.29  &  1.14  &  1.15   &   $\pm$ & 0.18   &   2.36  &  2.19    &   $\pm$ & 0.27      \\             
He~{\sc i} 5876    &    18.9   &  17.3   &   $\pm$ & 0.78  &     18.7   &  17.6   &   $\pm$ & 0.77  &  15.4  &  15.5   &   $\pm$ & 0.5    &   17.8  &  16.3    &   $\pm$ & 0.7      \\             
$\textrm{[OI]}$ 6300    &    8.63   &  7.68   &   $\pm$ & 0.46  &      7.90  &  7.32   &   $\pm$ & 0.37  &  6.23  &  6.27   &   $\pm$ & 0.31   &   10.8  &  9.67    &   $\pm$ & 0.47      \\             
$\textrm{[S~{\sc iii}]}$ 6312\tablefootmark{g}    &    2.43   &  2.15   &   $\pm$ & 0.33  &      2.17  &  2.00   &   $\pm$ & 0.19  &  2.20  &  2.17   &   $\pm$ & 0.19   &   1.93  &  1.68    &   $\pm$ & 0.15      \\      
$\textrm{[OI]}$ 6363    &    3.16   &  2.80   &   $\pm$ & 0.30  &      2.58  &  2.38   &   $\pm$ & 0.19  &  2.46  &  2.47   &   $\pm$ & 0.21   &   3.47  &  3.10    &   $\pm$ & 0.20      \\             
$\textrm{[N~{\sc ii}]}$ 6548    &    12.6   &  11.1   &   $\pm$ & 0.6   &     28.2   &  25.9   &   $\pm$ & 1.35  &  16.0  &  16.1   &   $\pm$ & 0.5    &   28.2  &  24.9    &   $\pm$ & 1.3      \\             
H$\alpha$ 6563\tablefootmark{h}    &    327    &  286    &   $\pm$ & 15    &    312.6   &  286.0  &   $\pm$ & 14.9  &  279   &  279    &   $\pm$ & 8      &   327   &  286     &   $\pm$ & 15      \\             
$\textrm{[N~{\sc ii}]}$ 6583    &    31.8   &  27.9   &   $\pm$ & 1.49  &     82.1   &  75.2   &   $\pm$ & 3.93  &  46.6  &  47.0   &   $\pm$ & 1.4    &   81.5  &  71.8    &   $\pm$ & 3.7      \\             
He~{\sc i} 6678\tablefootmark{i}    &    4.31   &  3.73   &   $\pm$ & 0.26  &      4.87  &  4.42   &   $\pm$ & 0.29  &  3.89  &  3.83   &   $\pm$ & 0.18   &   4.32  &  3.69    &   $\pm$ & 0.22      \\             
$\textrm{[S~{\sc ii}]}$ 6716    &    1.58   &  1.37   &   $\pm$ & 0.22  &      2.05  &  1.87   &   $\pm$ & 0.18  &  2.66  &  2.68   &   $\pm$ & 0.17   &   2.93  &  2.56    &   $\pm$ & 0.16      \\             
$\textrm{[S~{\sc ii}]}$ 6730    &    2.70   &  2.34   &   $\pm$ & 0.20  &      4.25  &  3.87   &   $\pm$ & 0.25  &  4.43  &  4.46   &   $\pm$ & 0.19   &   5.56  &  4.85    &   $\pm$ & 0.28      \\             
$\textrm{[Ar~{\sc v}]}$ 7005    &    --     &         &    --   &       &      --    &         &    --   &       &  --    &         &    --   &        &   --    &          &   --    &           \\             
He~{\sc i} 7065    &    13.4   &  11.4   &   $\pm$ & 0.7   &      10.9  &  9.8    &   $\pm$ & 0.60  &  9.11  &  9.16   &   $\pm$ & 0.33   &   11.0  &  9.39    &   $\pm$ & 0.57      \\             
$\textrm{[Ar~{\sc iii}]}$ 7136    &    11.5   &  9.8    &   $\pm$ & 0.6   &      18.0  &  16.1   &   $\pm$ & 0.99  &  10.3  &  10.3   &   $\pm$ & 0.3    &   17.6  &  15.0    &   $\pm$ & 0.92     \\             
He~{\sc i} 7281    &    1.02   &  0.86   &   $\pm$ & 0.14  &      0.67  &  0.60   &   $\pm$ & 0.14  &  1.46  &  1.47   &   $\pm$ & 0.20   &   0.96  &  0.81    &   $\pm$ & 0.10      \\             
$\textrm{[O~{\sc ii}]}$ 7320    &    8.24   &  6.90   &   $\pm$ & 0.48  &      4.94  &  4.39   &   $\pm$ & 0.31  &  4.00  &  4.02   &   $\pm$ & 0.18   &   7.53  &  6.35    &   $\pm$ & 0.42      \\             
$\textrm{[O~{\sc ii}]}$ 7330    &    8.06   &  6.75   &   $\pm$ & 0.47  &      4.72  &  4.20   &   $\pm$ & 0.32  &  4.15  &  4.17   &   $\pm$ & 0.21   &   6.69  &  5.64    &   $\pm$ & 0.37      \\                   
 $\textrm{[Ar~{\sc iii}]}$ 7751    &    2.37   &  1.94   &   $\pm$ & 0.25  &      2.75  &  2.40   &   $\pm$ & 0.22  &  1.92  &  1.93   &   $\pm$ & 0.18   &   3.34  &  2.74    &   $\pm$ & 0.21     \\ \hline      


\hline
  \multicolumn{1}{c}{}        
   &   \multicolumn{4}{c}{M1596}  &  \multicolumn{4}{c}{M2471} &   \multicolumn{4}{c}{M2860}    & \multicolumn{4}{c}{M1074}  &  \multicolumn{4}{c}{M1675} \\
 \hline

  \multicolumn{1}{c}{Ident.} &   \multicolumn{1}{c}{F}  & \multicolumn{3}{c}{I $\pm$ $\Delta$I}
  & \multicolumn{1}{c}{F} &  \multicolumn{3}{c}{ I $\pm$ $\Delta$I} &  \multicolumn{1}{c}{F} & \multicolumn{3}{c}{I $\pm$ $\Delta$I}   &  \multicolumn{1}{c}{F} & \multicolumn{3}{c}{I $\pm$ $\Delta$I}  &  \multicolumn{1}{c}{F} & \multicolumn{3}{c}{I $\pm$ $\Delta$I}  \\
 \hline

$\textrm{[O~{\sc ii}]}$ 3727     &     55.3   & 63.7     &   $\pm$ & 3.6   &  60.1   &  63.0   &   $\pm$ & 3.5   &  19.2   &  20.8   &   $\pm$ & 1.2   &  17.8   &  19.3   &  $\pm$  & 1.2   &   11.9   &  14.2   &   $\pm$  &  1.3   \\
H10 3798     &     --     &          &   --    &       &  4.19   &  4.39   &   $\pm$ & 0.59  &  2.93   &  3.15   &   $\pm$ & 0.69  &  --     &         &  --     &       &    --    &         &  --      &         \\
H9  3835     &     7.10   & 8.10     &   $\pm$ & 1.19  &  4.69   &  4.91   &   $\pm$ & 0.54  &  3.52   &  3.78   &   $\pm$ & 0.36  &  6.24   &  6.72   &   $\pm$ & 0.70  &   3.63   &  4.28   &   $\pm$  &  1.37   \\
$\textrm{[Ne~{\sc iii}]}$ 3868     &     110    & 125      &   $\pm$ & 6     &  111    &  116    &   $\pm$ & 6     &  80.4   &  86.1   &   $\pm$ & 4.4   &  103    &  110    &   $\pm$ &  6    &   78.0   &  91.6   &   $\pm$  &  4.9   \\
He~{\sc i}+H8  3889     &     10.1   & 11.5     &   $\pm$ & 1.0   &  16.2   &  16.9   &   $\pm$ & 1.0   &  12.0    &  12.9   &   $\pm$ & 0.7   &  15.6   &  16.7   &   $\pm$ & 1.0   &   10.7   &  12.5   &   $\pm$  &  1.9   \\
$\textrm{[Ne~{\sc iii}]}$ 3967\tablefootmark{a}      &     48. 7  & 37.7     &   $\pm$ & 2.8   &  48.9   &  34.0   &   $\pm$ & 2.5   &  45.1   &  30.7   &   $\pm$ & 2.4   &  46.4   &  32.2   &   $\pm$ & 2.5   &   35.6   &  24.0   &   $\pm$  &  2.6   \\    
He~{\sc i} 4026     &     4.93   & 5.52     &   $\pm$ & 1.42  &  2.21   &  2.30   &   $\pm$ & 0.76  &  3.35   &  3.56   &   $\pm$ & 0.51  &  1.42   &  1.51   &   $\pm$ & 0.48  &   --     &         &  --      &           \\
$\textrm{[S~{\sc ii}]}$ 4068     &     3.41   & 3.80     &   $\pm$ & 1.21  &  2.95   &  3.07   &   $\pm$ & 0.50  &  2.58   &  2.72   &   $\pm$ & 1.05 &  1.86   &  1.97   &   $\pm$ & 0.37  &   --     &         &  --      &           \\
$\textrm{[S~{\sc ii}]}$ 4071     &    --      &          &   --    &       &  --     &         &   --    &       &  1.40   &  1.48   &   $\pm$ & 0.74 &    --     &         &   --    &       &    --      &         &     --     &        \\
H6 4101\tablefootmark{b}    &     24.6   & 26.9     &   $\pm$ & 1.7   &  26.1   &  26.7   &   $\pm$ & 1.3   &  24.7   &  26.0   &   $\pm$ & 1.21  &  23.3   &  24.6   &   $\pm$ & 1.2   &   20.4   &  23.0   &   $\pm$  &  2.1   \\
H5 4340\tablefootmark{c}    &     43.8   & 46.5     &   $\pm$ & 2.1   &  44.9   &  45.6   &   $\pm$ & 1.8   &  42.6   &  44.1   &   $\pm$ & 1.70  &  42.8   &  44.5   &   $\pm$ & 1.7   &   42.6   &  46.3   &   $\pm$  &  2.1   \\
$\textrm{[O~{\sc iii}]}$ 4363     &     18.5   & 19.9     &   $\pm$ & 1.4   &  16.5   &  17.0   &   $\pm$ & 0.8   &  9.61   &  9.97   &   $\pm$ & 0.54  &  16.3   &  16.9   &   $\pm$ & 0.8   &   11.5   &  12.5   &   $\pm$  &  1.6   \\
$\textrm{Ne~{\sc ii}}$ 4428     &            &          &         &       &         &         &         &       &  --     &         &  --     &       &  --     &         &    --   &       &    --    &         &    --    &         \\
He~{\sc i} 4472     &     4.05   & 4.31     &   $\pm$ & 0.51  &  3.71   &  3.81   &   $\pm$ & 0.32  &  5.28   &  5.44   &   $\pm$ & 0.39  &  5.12   &  5.27   &   $\pm$ & 0.59  &   4.05   &  4.33   &   $\pm$  &  0.84    \\
He~{\sc ii} 4542     &     --     &          &   --    &       &   --    &         &   --    &       &  --     &         &   --    &       &   --    &         &    --   &       &    --    &         &    --    &          \\
N~{\sc iii}+O~{\sc ii} 4640     &     4.44   & 4.63     &   $\pm$ & 0.59  &  3.54   &  3.62   &   $\pm$ & 0.52  &  3.00   &  3.05   &   $\pm$ & 0.54  &   --    &         &    --   &       &    --    &         &    --    &           \\ 
O~{\sc ii} 4650     &     --     &          &   --    &       &  --     &         &  --     &       &     --    &         &    --     &       &   --    &         &    --   &       &    --    &         &    --    &          \\
He~{\sc ii} 4686     &     34.0   & 35.3     &   $\pm$ & 1.2   & 31.8    &  32.5   &   $\pm$ & 1.0   &  7.24   &  7.35   &   $\pm$ & 0.37  &  2.59   &  2.62   &   $\pm$ & 0.38  &   12.0   &  12.4   &   $\pm$  &  0.7   \\
$\textrm{[Ar~{\sc iv}]}$ 4711\tablefootmark{d}    &     6.86   & 6.42     &   $\pm$ & 0.53  & 4.53    &  4.06   &   $\pm$ & 0.40  &  3.48   &  2.79   &   $\pm$ & 0.31  &  2.38   &  1.58   &   $\pm$ & 0.29  &   3.66   &  3.06   &   $\pm$  &  0.69    \\
$\textrm{[Ar~{\sc iv}]}$  4740     &     7.80   & 8.04     &   $\pm$ & 0.50  & 4.58    &  4.67   &   $\pm$ & 0.34  &  4.27   &  4.32   &   $\pm$ & 0.38  &  3.74   &  3.78   &   $\pm$ & 0.27  &   4.25   &  4.35   &   $\pm$  &  0.69    \\
H$\beta$ 4861\tablefootmark{e}    &     100    & 100      &   $\pm$ & 3     & 100     &  100    &   $\pm$ & 3     &  100    &  100    &   $\pm$ & 3     & 100     &  100    &   $\pm$ & 3     &   100    &  100    &   $\pm$  &  3      \\
He~{\sc i} 4922     &     --       &          &    --     &       & 0.74    &  0.75   &   $\pm$ & 0.25  &  1.36   &  1.36   &   $\pm$ & 0.85  &  --     &         &  --     &       &   2.43   &  2.42   &   $\pm$  &  0.71    \\
$\textrm{[O~{\sc iii}]}$ 4959     &     582    &  587     &   $\pm$ & 17    &  510    &  517    &   $\pm$ & 15    &  493    &  492    &   $\pm$ & 15    &  482    &  479    &   $\pm$ & 15    &   527    &  523    &   $\pm$  &  16     \\ 
$\textrm{[O~{\sc iii}]}$ 5007      &    1747    &  1732    &   $\pm$ & 52    & 1536    & 1554    &   $\pm$ & 47   & 1479    & 1472    &   $\pm$ & 45    & 1456    & 1446    &   $\pm$ & 44    &  1609    & 1587    &   $\pm$  &  48     \\
$\textrm{[NI]}$ 5198     &     --     &          &   --    &       &  --     &         &    --   &       &  --     &         &   --    &       &  --     &         &  --     &       &   2.28   &  2.20   &   $\pm$  &  0.70    \\
He~{\sc ii} 5411     &     3.23   &  3.13    &   $\pm$ & 0.38  & 2.86    &  2.87   &   $\pm$ & 0.19  &  0.62   &  0.61   &   $\pm$ & 0.19  &  --     &         &  --     &       &   --     &         &   --     &           \\
$\textrm{[Cl~{\sc iii}]}$ 5518     &     0.67   &  0.64    &   $\pm$ & 0.35  &  --     &         &    --   &       &  0.33   &  0.32   &   $\pm$ & 0.12  &  --     &         &  --     &       &   --     &         &   --     &          \\
$\textrm{[Cl~{\sc iii}]}$ 5538     &     1.33   &  1.28    &   $\pm$ & 0.34  &  --     &         &    --   &       &  0.50   &  0.49   &   $\pm$ & 0.17  &  --     &         &  --     &       &   --     &         &   --     &          \\
$\textrm{[N~{\sc ii}]}$ 5755     &     2.74   &  2.60    &   $\pm$ & 0.40  & 1.08    &  1.08   &   $\pm$ & 0.17  &  1.00   &  0.96   &   $\pm$ & 0.18  &  0.54   &  0.52   &   $\pm$ & 0.20  &   2.95   &  2.69   &   $\pm$  &  0.50     \\
He~{\sc i} 5876     &     15.4   &  14.5    &   $\pm$ & 0.7   & 12.8    &  12.8   &   $\pm$ & 0.6   &  17.3   &  16.6   &   $\pm$ & 0.7   &  17.8   &  16.9   &   $\pm$ & 0.7   &  17.9    &  16.2   &   $\pm$  &  0.9     \\
$\textrm{[OI]}$ 6300     &     7.83   &  7.20    &   $\pm$ & 0.38  & 8.16    &  8.07   &   $\pm$ & 0.42  &  3.02   &  2.86   &   $\pm$ & 0.19  &  4.08   &  3.83   &   $\pm$ & 0.26  &   10.7  &  9.3     &   $\pm$  &  0.7      \\
$\textrm{[S~{\sc iii}]}$ 6312 \tablefootmark{g}     &     4.69   &  4.18    &   $\pm$ & 0.29  & 1.74    &  1.61   &   $\pm$ & 0.24  &  1.88   &  1.75   &   $\pm$ & 0.17  &  2.38   &  2.23   &   $\pm$ & 0.22  &   3.17   &  2.73   &   $\pm$  &  0.56     \\
$\textrm{[OI]}$ 6363     &     2.44   &  2.24    &   $\pm$ & 0.18  & 2.53    &  2.50   &   $\pm$ & 0.24  &  1.27   &  1.20   &   $\pm$ & 0.18  &  1.35   &  1.27   &   $\pm$ & 0.22  &   3.32   &  2.89   &   $\pm$  &  0.53     \\
$\textrm{[N~{\sc ii}]}$ 6548     &     41.5   &  37.7    &   $\pm$ & 1.9   & 18.6    &  18.3   &   $\pm$ & 1.0   &  13.0   &  12.2   &   $\pm$ & 0.6   &  6.84   &  6.38   &   $\pm$ & 0.38  &   41.8   &  36.0   &   $\pm$  &  1.9    \\
H$\alpha$ 6563\tablefootmark{h}     &     321    &  286     &   $\pm$ & 15    & 295     &  286    &   $\pm$ & 15    &  306    &  286    &   $\pm$ & 15    &  307    &  286    &   $\pm$ & 15    &  335     &  286    &   $\pm$  &  15    \\
$\textrm{[N~{\sc ii}]}$ 6583     &     121    &  110     &   $\pm$ & 6     & 53.7    &  52.9   &   $\pm$ & 2.7   &  35.9   &  33.7   &   $\pm$ & 1.8   &  19.8   &  18.5   &   $\pm$ & 1.0   &  123     &  105    &   $\pm$  &  6      \\
He~{\sc i} 6678\tablefootmark{i}    &     4.66   &  3.96    &   $\pm$ & 0.31  & 3.06    &  2.79   &   $\pm$ & 0.26  &  4.29   &  3.97   &   $\pm$ & 0.24  &  4.05   &  3.74   &   $\pm$ & 0.25  &   6.36   &  5.33   &   $\pm$  &  0.45     \\
$\textrm{[S~{\sc ii}]}$ 6716     &     7.19   &  6.47    &   $\pm$ & 0.39  & 3.37    &  3.32   &   $\pm$ & 0.28  &  2.28   &  2.13   &   $\pm$ & 0.17  &  1.42   &  1.32   &   $\pm$ & 0.18  &   5.03   &  4.27   &   $\pm$  &  0.39     \\
$\textrm{[S~{\sc ii}]}$ 6730     &     12.2   &  10.9    &   $\pm$ & 0.6   & 5.60    &  5.51   &   $\pm$ & 0.35  &  3.80   &  3.55   &   $\pm$ & 0.23  &  2.24   &  2.08   &   $\pm$ & 0.20  &   10.6   &  9.01   &   $\pm$  &  0.62     \\
$\textrm{[Ar~{\sc v}]}$ 7005     &     1.23   &  1.09    &   $\pm$ & 0.14  &  0.58   &  0.57   &   $\pm$ & 0.20  &  --     &         &   --    &       &  --     &         &   --    &       &  --      &         &   --     &           \\
He~{\sc i} 7065     &     7.94   &  7.02    &   $\pm$ & 0.45  &  6.92   &  6.78   &   $\pm$ & 0.44  &  7.96   &  7.37   &   $\pm$ & 0.47  &  10.1   &  9.2    &   $\pm$ & 0.6   &  8.57    &  7.10   &   $\pm$  &  0.49     \\
$\textrm{[Ar~{\sc iii}]}$ 7136     &     26.5   &  23.3    &   $\pm$ & 1.4   &  12.8   &  12.5   &   $\pm$ & 0.8   &  15.7   &  14.5   &   $\pm$ & 0.9   &  9.07   &  8.30   &   $\pm$ & 0.53  &  17.9    &  14.8   &   $\pm$  &  1.0     \\
He~{\sc i} 7281     &     --     &          &   --    &       &  0.71   &  0.69   &   $\pm$ & 0.19  &  0.41   &  0.38   &   $\pm$ & 0.16  &  --     &         &  --     &       &   --     &         &  --      &           \\
$\textrm{[O~{\sc ii}]}$ 7320     &     5.21   &  4.55    &   $\pm$ & 0.34  & 4.60    &  4.49   &   $\pm$ & 0.35  &  2.42   &  2.22   &   $\pm$ & 0.21  &  4.45   &  4.05   &   $\pm$ & 0.31  &   --     &         &  --      &           \\
$\textrm{[O~{\sc ii}]}$ 7330     &     4.79   &  4.18    &   $\pm$ & 0.33  & 4.24    &  4.14   &   $\pm$ & 0.36  &  2.29   &  2.10   &   $\pm$ & 0.22  &  4.55   &  4.14   &   $\pm$ & 0.33  &   4.20   &  3.42   &   $\pm$  &  0.38     \\
$\textrm{[Ar~{\sc iii}]}$ 7751     &     5.13   &  4.38    &   $\pm$ & 0.35  & 2.85    &  2.77   &   $\pm$ & 0.28  &  3.11   &  2.82   &   $\pm$ & 0.25  &  1.80   &  1.61   &   $\pm$ & 0.18  &   3.81   &  3.10   &   $\pm$  &  0.35   \\ 
\hline
\end{longtable}         

\begin{flushleft}
$^{a}$ Corrected for the flux contribution of H~{\sc I} $\lambda$3970.07 and He~{\sc i} $\lambda$3964.73. \\
$^{b}$ Corrected for the flux contribution of He~{\sc ii} $\lambda$4100.04. \\                   
$^{c}$ Corrected for the flux contribution of He~{\sc ii} $\lambda$4338.67. \\                
$^{d}$ Corrected for the flux contribution of He~{\sc i} $\lambda$4713.14. \\                    
$^{e}$ Corrected for the flux contribution of He~{\sc ii} $\lambda$4859.32 \\                   
$^{f}$ Saturated. Flux obtained from the theoretical relation between [O~{\sc iii}] $\lambda5007$ and $\lambda4959$.  \\                                    $^{g}$ Corrected for the flux contribution of He~{\sc ii} $\lambda$6310.85. \\                   
$^{h}$ Corrected for the flux contribution of He~{\sc ii} $\lambda$6560.10. \\                   
$^{i}$ Corrected for the flux contribution of He~{\sc ii} $\lambda$6683.20. \\                    
\end{flushleft}

\clearpage

\begin{longtable}{l@{\hspace{8pt}} c@{\hspace{8pt}} c@{\hspace{8pt}} c@{\hspace{8pt}} c c@{\hspace{8pt}}}
\caption{Ionic abundances.}          
\label{table:ionic_abundances}  \\
\hline 
\multicolumn{1}{c}{}&\multicolumn{1}{c}{M1687*}    & \multicolumn{1}{c}{M2068*}  &  \multicolumn{1}{c}{M2538*} &  \multicolumn{1}{c}{M50*} \tabularnewline
\hline 
\multicolumn{1}{c}{Ion} & 
\multicolumn{4}{c}{X$^{+i}$/H$^+$} \\
\hline 
He$^{+}$(5876)   &  0.094($\pm$0.005)                   & 0.106($\pm$0.005)                   & 9.26($\pm$0.40)$\times$$10^{-02}$   & 9.78($\pm$0.45)$\times$$10^{-02}$ \\ [3pt]
He$^{2+}$(4686)  &  2.50($\pm$0.43)$\times$$10^{-03}$   & 3.20($\pm$0.22)$\times$$10^{-03}$   & 1.02($\pm$0.04)$\times$$10^{-02}$   & 1.21($\pm$0.05)$\times$$10^{-02}$\\ [3pt]
O$^{+}$(3727)    &  4.94($\pm$ 2.33)$\times$$10^{-06}$  &  8.19($\pm_{8.19}^{9.58}$)$\times$$10^{-05}$   & 1.39 ($\pm$0.62)$\times$$10^{-05}$  & 1.39($\pm$2.20)$\times$$10^{-05}$\\
O$^{+}$(7320)    &  2.72($\pm$0.78)$\times$$10^{-05}$   & 3.00($\pm$2.37)$\times$$10^{-05}$     & 2.44($\pm$1.55)$\times$$10^{-05}$   & 2.21($\pm$1.33)$\times$$10^{-05}$\\
O$^{+}$(7330)    &  3.12($\pm$0.90)$\times$$10^{-05}$   & 3.55($\pm$2.77)$\times$$10^{-05}$    & 2.97($\pm$1.90)$\times$$10^{-05}$   & 2.29($\pm$1.40)$\times$$10^{-05}$\\
O$^{+}$ (Adopted)          &  2.11($\pm$0.50)$\times$$10^{-05}$   & 4.91$\pm$3.18)$\times$$10^{-05}$   & 2.27($\pm$1.00)$\times$$10^{-05}$   & 1.96($\pm$0.80)$\times$$10^{-05}$  \\ [3pt]
O$^{+2}$(4959)   &  3.05($\pm$0.34)$\times$$10^{-04}$   & 4.53($\pm$0.49)$\times$$10^{-04}$   & 2.68($\pm$0.23)$\times$$10^{-04}$   & 3.71($\pm$0.32)$\times$$10^{-04}$\\ [3pt]
Ne$^{+2}$(3869)  &  7.46($\pm$0.87)$\times$$10^{-05}$   & 1.17($\pm$0.13)$\times$$10^{-04}$  & 6.44($\pm$1.42)$\times$$10^{-06}$   & 7.67($\pm$0.69)$\times$$10^{-05}$\\ [3pt]
Ar$^{+2}$(7135)  &  4.79($\pm$0.41)$\times$$10^{-07}$   & 1.15($\pm$0.01)$\times$$10^{-06}$   & 5.60($\pm$0.31)$\times$$10^{-07}$   & 9.15($\pm$0.67)$\times$$10^{-07}$\\ [3pt]
Ar$^{+3}$(4711)  &  5.58($\pm$0.64)$\times$$10^{-07}$   & 7.27($\pm$0.87)$\times$$10^{-07}$   & 4.83($\pm$0.53)$\times$$10^{-07}$   & 6.78($\pm$0.63)$\times$$10^{-07}$\\
Ar$^{+3}$(4740)  &  5.58($\pm$0.63)$\times$$10^{-07}$   & 7.26($\pm$0.87)$\times$$10^{-07}$   & 4.82($\pm$0.53)$\times$$10^{-07}$   & 6.78($\pm$0.63)$\times$$10^{-07}$\\
Ar$^{+3}$ (Adopted)          &  5.58($\pm$0.60)$\times$$10^{-07}$   & 7.27($\pm$0.80)$\times$$10^{-07}$   & 4.82($\pm$0.52)$\times$$10^{-07}$   & 6.78($\pm$0.63)$\times$$10^{-07}$ \\ [3pt]
Ar$^{+4}$(7005)  &         ---                          &        ---                          &       ---       &        ---                  \\ [3pt]
N$^{+}$(6584)    &  3.34($\pm$0.35)$\times$$10^{-06}$   & 1.98($\pm$0.67 )$\times$$10^{-05}$    & 6.50($\pm$1.48)$\times$$10^{-06}$   & 8.71($\pm$3.22)$\times$$10^{-06}$\\ [3pt]
S$^{+}$(6716)    &  1.09($\pm$0.52)$\times$$10^{-07}$   & 8.41($\pm$4.51 )$\times$$10^{-07}$    & 2.04($\pm$0.60)$\times$$10^{-07}$   & 2.51($\pm$2.50)$\times$$10^{-07}$ \\
S$^{+}$(6731)    &  1.05($\pm$0.52)$\times$$10^{-07}$   & 8.06($\pm$4.21 )$\times$$10^{-07}$    & 2.04($\pm$0.60)$\times$$10^{-07}$   & 2.51($\pm$2.50)$\times$$10^{-07}$\\
S$^{+}$ (Adopted)          &  1.07($\pm$0.41)$\times$$10^{-07}$   & 8.24($\pm$3.40 )$\times$$10^{-07}$    & 2.04($\pm$0.46)$\times$$10^{-07}$   & 2.51($\pm$1.13)$\times$$10^{-07}$  \\ [3pt]
S$^{+2}$(6312)   &  1.74($\pm$0.32)$\times$$10^{-06}$   & 3.13($\pm$0.42)$\times$$10^{-06}$   & 2.19($\pm$0.26)$\times$$10^{-06}$   & 2.04($\pm$0.24)$\times$$10^{-06}$\\ [3pt]
Cl$^{+2}$(5517)  &         ---                          &        ---                          &         ---                    & 5.86($\pm$3.03)$\times$$10^{-08}$\\
Cl$^{+2}$(5537)  &          ---                         &        ---                          &         ---       & 7.07($\pm$1.68)$\times$$10^{-08}$\\
Cl$^{+2}$ (Adopted)          &          ---                            &                ---                     &     ---      & 6.47($\pm$1.86)$\times$$10^{-08}$ \\
\hline 
\multicolumn{1}{c}{}  & \multicolumn{1}{c}{M1596}    & \multicolumn{1}{c}{M2471}  &  \multicolumn{1}{c}{M2860} &  \multicolumn{1}{c}{M1074} &  \multicolumn{1}{c}{M1675}\tabularnewline
\hline 
\multicolumn{1}{c}{Ion} &  
\multicolumn{5}{c}{X$^{+i}$/H$^+$} \\
\hline
He$^{+}$(5876)  & 8.86($\pm$0.47)$\times$$10^{-02}$   & 7.95($\pm$0.42)$\times$$10^{-02}$  & 0.106($\pm$0.005)                     & 9.82($\pm$0.45)$\times$$10^{-02}$  & 1.01($\pm$0.06)$\times$$10^{-01}$      \\ [3pt]
He$^{2+}$(4686) & 2.99($\pm$0.11)$\times$$10^{-02}$   & 2.74($\pm$0.09)$\times$$10^{-02}$  & 6.09 ($\pm$0.32)$\times$$10^{-03}$    & 2.24($\pm$0.31)$\times$$10^{-03}$  & 1.03($\pm$0.06)$\times$$10^{-02}$      \\ [3pt]
O$^{+}$(3727)   & 1.92($\pm_{1.91}^{2.28}$)$\times$$10^{-05}$      & 2.41($\pm$2.39)$\times$$10^{-05}$     & 4.75($\pm$3.67)$\times$$10^{-06}$     & 5.19($\pm$2.60)$\times$$10^{-06}$  & 7.03($\pm_{7.00}^{10.96}$)$\times$$10^{-06}$        \\
O$^{+}$(7320)   & 2.99($\pm$1.72)$\times$$10^{-05}$   & 4.41($\pm$2.58)$\times$$10^{-05}$  & 1.04($\pm$0.84)$\times$$10^{-05}$     & 2.76($\pm$0.74)$\times$$10^{-05}$  & 1.78($\pm$ 1.55)$\times$$10^{-05}$        \\
O$^{+}$(7330)   & 3.23($\pm$1.83)$\times$$10^{-05}$   & 4.77($\pm$2.77)$\times$$10^{-05}$  & 1.15($\pm$0.92)$\times$$10^{-05}$     & 3.31($\pm$0.90)$\times$$10^{-05}$  & 2.00($\pm$ 1.83)$\times$$10^{-05}$        \\
O$^{+}$ (Adopted)         & 2.71($\pm$1.18)$\times$$10^{-05}$   & 3.86($\pm$1.50)$\times$$10^{-05}$  & 8.87($\pm$4.09)$\times$$10^{-06}$     & 2.20($\pm$0.50)$\times$$10^{-05}$  & 1.50($\pm$1.25)$\times$$10^{-05}$        \\ [3pt]
O$^{+2}$(4959)  & 3.48($\pm$0.37)$\times$$10^{-04}$   & 3.16($\pm$0.26)$\times$$10^{-04}$  & 5.11($\pm$0.44)$\times$$10^{-04}$     & 2.99($\pm$0.26)$\times$$10^{-04}$  & 4.54($\pm$0.76)$\times$$10^{-04}$      \\ [3pt]
Ne$^{+2}$(3869) & 6.99($\pm$0.82)$\times$$10^{-05}$   & 6.73($\pm$0.60)$\times$$10^{-05}$  & 9.35($\pm$0.88)$\times$$10^{-05}$     & 6.35($\pm$0.59)$\times$$10^{-05}$  & 8.04($\pm$1.59)$\times$$10^{-05}$      \\ [3pt]
Ar$^{+2}$(7135) & 1.31($\pm$0.12)$\times$$10^{-06}$   & 7.18($\pm$0.56)$\times$$10^{-07}$  & 1.20($\pm$0.09)$\times$$10^{-06}$     & 4.75($\pm$0.37)$\times$$10^{-07}$  & 1.08($\pm$0.13)$\times$$10^{-06}$      \\ [3pt]
Ar$^{+3}$(4711) & 1.24($\pm$0.13)$\times$$10^{-06}$   & 7.67($\pm$0.66)$\times$$10^{-07}$  & 1.10($\pm$0.11)$\times$$10^{-06}$     & 5.08($\pm$0.49)$\times$$10^{-07}$  & 9.49($\pm$1.95)$\times$$10^{-07}$     \\
Ar$^{+3}$(4740) & 1.24($\pm$0.13)$\times$$10^{-06}$   & 7.67($\pm$0.66)$\times$$10^{-07}$  & 1.09($\pm$0.11)$\times$$10^{-06}$     & 5.08($\pm$0.49)$\times$$10^{-07}$  & 9.49($\pm$1.95)$\times$$10^{-07}$     \\
Ar$^{+3}$ (Adopted)         & 1.24($\pm$0.14)$\times$$10^{-06}$   & 7.67($\pm$0.63)$\times$$10^{-07}$  & 1.09($\pm$0.11)$\times$$10^{-06}$     & 5.08($\pm$0.50)$\times$$10^{-07}$  & 9.49($\pm$1.84)$\times$$10^{-07}$     \\ [3pt]
Ar$^{+4}$(7005) & 1.34($\pm$0.20)$\times$$10^{-07}$   & 7.19($\pm$2.56)$\times$$10^{-08}$  & ---                                   & ---                                & ---                                   \\ [3pt]
N$^{+}$(6584)   & 1.58($\pm$0.36)$\times$$10^{-05}$   & 8.96($\pm$2.30)$\times$$10^{-06}$  & 4.11($\pm$1.15)$\times$$10^{-06}$     & 2.58($\pm$0.23)$\times$$10^{-06}$  & 1.68($\pm$0.61)$\times$$10^{-05}$     \\ [3pt]
S$^{+}$(6716)   & 5.30($\pm$1.90)$\times$$10^{-07}$   & 3.02($\pm$2.04)$\times$$10^{-07}$  & 1.46($\pm$0.73)$\times$$10^{-07}$     & 9.26($\pm$4.74)$\times$$10^{-08}$  & 5.23($\pm$4.56)$\times$$10^{-07}$     \\
S$^{+}$(6731)   & 5.30($\pm$1.90)$\times$$10^{-07}$   & 3.02($\pm$2.04)$\times$$10^{-07}$  & 1.46($\pm$0.73)$\times$$10^{-07}$     & 9.19($\pm$4.73)$\times$$10^{-08}$  & 5.88($\pm$4.35)$\times$$10^{-07}$     \\ 
S$^{+}$ (Adopted)         & 5.30($\pm$1.90)$\times$$10^{-07}$   & 3.02($\pm$0.90)$\times$$10^{-07}$  & 1.46($\pm$0.40)$\times$$10^{-07}$     & 9.23($\pm$2.12)$\times$$10^{-08}$  & 5.56($\pm$3.26)$\times$$10^{-07}$  \\ [3pt]
S$^{+2}$(6312)  & 4.50($\pm$0.58)$\times$$10^{-06}$   & 1.80($\pm$0.29)$\times$$10^{-06}$  & 3.71($\pm$0.48)$\times$$10^{-06}$     & 2.38($\pm$0.30)$\times$$10^{-06}$  & 4.64($\pm$1.35)$\times$$10^{-06}$     \\ [3pt]
Cl$^{+2}$(5517) & 6.46($\pm$3.46)$\times$$10^{-08}$   & ---                                & 6.12($\pm$2.42)$\times$$10^{-08}$     & ---                                & ---                                     \\        
Cl$^{+2}$(5537) & 9.94($\pm$2.76)$\times$$10^{-08}$   & ---                                & 6.16($\pm$2.17)$\times$$10^{-08}$     & ---                                & ---                                      \\     
Cl$^{+2}$ (Adopted)        & 8.20($\pm$2.10)$\times$$10^{-08}$   & ---                                & 6.14($\pm$1.55)$\times$$10^{-08}$     & ---                                & ---    \\
\hline
\end{longtable}

\end{appendix}

\end{document}